\documentclass[aps,prd,reprint,longbibliography,titlepage,nofootinbib]{revtex4-1}

\usepackage{amsthm}
\usepackage{amssymb}   
\usepackage{dsfont}
\usepackage{mathtools} 
\usepackage{hyperref}
\hypersetup{colorlinks=true,linkcolor=blue,urlcolor=blue,citecolor=blue}
\usepackage{accents}
\usepackage{tensor}
\usepackage{xcolor}
\usepackage{graphicx}
\usepackage[cal=boondox]{mathalfa}
\usepackage{lipsum}
\usepackage{relsize}
\usepackage{color}
\usepackage{comment}
\usepackage{nameref}
\usepackage{orcidlink}

\begin{document}

\title{Post-Newtonian Gravitational Waves with cosmological constant $\Lambda$ from the Einstein-Hilbert theory}

\author{Ricardo Escobedo\orcidlink{0000-0001-5815-4748}}
\email[Corresponding author \\]{ricardo.escobedo@academicos.udg.mx}

\author{Claudia Moreno\orcidlink{0000-0002-0496-032X}}
\email[]{claudia.moreno@academico.udg.mx}

\author{Rafael Hern\'andez-Jim\'enez\orcidlink{0000-0002-2740-9610}}
\email[]{rafael.hjimenez@academicos.udg.mx}
\affiliation{Departamento de F\'{i}sica, CUCEI, Universidad de Guadalajara, 44430, Guadalajara, Jalisco, M\'exico}
	
\date{\today}
	
\begin{abstract}
We study the Post-Newtonian approach implemented to the Einstein-Hilbert action adding the cosmological constant $\Lambda$ at 1PN order. We consider very small values of $\Lambda$ to derive the Lagrangian of a two body compact system at the center of mass frame at 1PN. Furthermore, the phase function $\phi(t)$ is obtained from the balance equation and the two polarizations $h_{+}$ and $h_{\times}$ are also calculated. We observe changes due to $\Lambda$ only at very low frequencies and we notice that it plays the role of ``stretch" the spacetime such that both amplitudes become smaller; however, given its nearly negligible value, $\Lambda$ has no relevance at higher frequencies whatsoever. 
%
\end{abstract}
	
\maketitle
	
	
\section{Introduction}\label{intro}
The direct detection of Gravitational Waves (GWs) has opened new perspectives to understand the nature and behavior of the universe from an astrophysical point of view~\cite{PhysRevLett.116.061102,PhysRevLett.119.161101}. These observations strengthen the General Relativity (GR) predictions given by Einstein in 1916~\cite{Einstein:1916cc}. On the other hand, from a plethora of gravitational phenomena, the very small value of the astrophysical cosmological constant $\Lambda \simeq10^{-52}\mathrm{m}^{-2}$~\cite{Planck:2018nkj} is probably the reason for not considering its contribution into the Einstein Field Equations (EFE); however, there are in fact astronomical observations that suggest that $\Lambda$ might cause the current accelerated expansion of the universe~\cite{Riess_1998, Riess_2000,Riess_2004,Eisenstein_2005}. For instance, the analysis of the cosmological microwave background radiation~\cite{Bernaris,Komatsu_2011} must include the effect of $\Lambda$; nonetheless, studies of probable observational effects inside the solar system, due to the $\Lambda$, are nearly negligible to be detected~\cite{PhysRevD.73.063004,Balaguera-Antolínez_2006}. Certainly, the standard cosmological $\Lambda$ Cold Dark Matter $(\Lambda \mathrm{CDM})$ model has been successfully tested throughout several sources of observations, and it remains the most simple yet accurate scenario; however, there are still areas of unresolved phenomenology and ignorance. 

Moreover, the Post-Newtonian (PN) expansion is implemented in GR to obtain approximate solutions of the EFE. This method consists in expanding the metric at various orders around small values of the velocity ratio $v/c$, where $v$ is the GWs velocity~\cite{Sergei_M_Kopeikin_2004}. Here, the Newtonian theory is recovered when taking the limit of the speed of light to infinity, or the velocity ratio to zero. Einstein first made use of the PN approximation (at first order) to compute the perihelion precession of the Mercury's orbit~\cite{Perihelum_Mercury}. Nowadays this method is mostly utilized to study the propagation of GWs of the relativistic two body problem (see for instance~\cite{PhysRevLett.14.241}). Note that this approach is only valid at the very near zone of the source $\mathcal{R}$; namely, in the region when the evaluation point $r$ ($0<r\ll \mathcal{R}$) is much smaller than the emitted wavelength $\lambda$; in other words the condition $r\ll\lambda$ must be satisfied; where in this region there is no retarded time~\cite{Maggiore:2007ulw,poisson2014gravity}. On the other hand, in the external domain $d<r<\infty$ we introduce the Post-Minkowskian (PM) approximation, with $d$ as the radius of the source; and here the gothic metric $\mathfrak{g}^{\mu\nu}:=\sqrt{-g}g^{\mu\nu}$ is written as an expansion of powers of the Newtonian constant of gravitation $G$. Note that there is an overlapping region $d<r<\mathcal{R}$; therefore, the coefficients of the faraway zone can be expressed in terms of powers of the PN approximation by matching relations~\cite{Maggiore:2007ulw}. The DIRE (Direct Integration of Relaxed Einstein Equation) can be used to compute a wave equation of the EFE in an exact form as long as the harmonic gauge holds and obtain waveforms as powers of PN orders~\cite{1975ApJ...197..717E, PhysRevD.54.4813}. 

The starting point is the EH action for the gravitational field with cosmological constant $\Lambda$ given by:
\begin{equation}
 S[g]=\int_M d^4 x\sqrt{-g}\bigg[ \frac{16\pi G}{c^4}\left(R-2\Lambda  \right)\bigg] +L_m \,,  
\end{equation}
where $G$ is the Newton's gravitational constant, $c$ is the speed of light in the vacuum, the metric determinant is $(-g):=\mathrm{det}(g_{\mu\nu})$, $L_m$ stands as the action that describes matter, $R:=g^{\mu\nu}R_{\mu\nu}$ and $R_{\mu\nu}:=R^{\alpha}{}_{\mu\nu\alpha}$ are the scalar curvature and Ricci curvature tensor, respectively. They are derived from the curvature tensor $R^{\alpha}{}_{\beta\gamma\delta}=\partial_{\gamma}\Gamma^{\alpha}{}_{\beta\delta}-\partial_{\delta}\Gamma^{\alpha}{}_{\beta\gamma}+\Gamma^{\rho}{}_{\beta\delta}\Gamma^{\alpha}{}_{\rho\gamma}-\Gamma^{\rho}{}_{\beta\gamma}\Gamma^{\alpha}{}_{\rho\delta}$, where the Christoffel symbols are in terms of the metric tensor and its partial derivatives $\Gamma^{\rho}{}_{\alpha\beta}=\frac{1}{2}g^{\rho\gamma}(\partial_{\alpha}g_{\gamma\beta}+\partial_{\beta}g_{\gamma\alpha}-\partial_{\gamma}g_{\alpha\beta})$. Thus, the EFE are given by:
\begin{equation}\label{EFE}
R_{\mu\nu}-\frac{1}{2}R g_{\mu\nu}+\Lambda g_{\mu\nu}=\frac{8\pi G}{c^4}T_{\mu\nu} \,,
\end{equation}
where the source term is obtained by $T^{\mu\nu}:=-\frac{2}{\sqrt{-g}}\frac{\partial \sqrt{-g} L_m}{\partial g_{\mu\nu}}$.

The main aim of this paper consists in exploring the effects on the propagation of GWs due to the presence of $\Lambda$, having a two body problem examined with a 1PN method. We then compare our results with those with $\Lambda=0$ at 2PN~\cite{PhysRevD.51.5360, Luc_Blanchet}, having the same inspiralling compact binaries system. Furthermore, previous studies have been explored the effects of $\Lambda$ in the linearized GR, the authors expanded the metric around a flat Minkowski space-time~\cite{PhysRevD.81.084002,PhysRevD.84.063523}; however, in our study the outcome of $\Lambda$ might have a rather geometrical description. 

We begin our analysis in Sec.~\ref{Sec.REFE} where we solve the EFE through the DIRE approach utilizing the gothic metric and imposing the harmonic gauge, we then compute the tensor waveforms $h^{ij}$ at the near zone contribution of the faraway components through the Epstein Wagoner (EW) tensors at 1PN order at the center of mass frame of a binary compact system taking a `Coulomb-like' gauge~\cite{10.1093/oso/9780198786399.001.0001}. Many explicit calculations are presented in Appendices~\ref{Metric_Newtonian_order} and~\ref{two_lagrangian}.  Then, in Sec.~\ref{Sec_Circ_orbit} the circular orbit properties are explained, the PN parameters $\gamma$ and $x$ are introduced, and we calculate the energy loss rate. After that, by taking into account the balance equation~\eqref{balance_equation} we obtain the propagation phase $\phi(t)$ of a gravitational wave at 1PN approximation. Note that these results can also be derived using the symmetric trace-free tensor (see Appendix \ref{STF}). Moreover, in Sec.~\ref{GW-hplus_hcross} we compute the polarizations waveforms $h_{+}(t)$ and $h_{\times}(t)$ and we present our results. We finish the paper by making some remarks in Sec.~\ref{Sec_conclusions}. 

\textit{Conventions.} We consider a 4-dimensional spacetime manifold $M$. Spacetime indices are designated by greek letters $\mu,\nu, \dots=\lbrace 0,i \rbrace$ where $i$ labels spatial components of tensors, whereas $0$ indicates the temporal component. These indices are raised and lowered with the spacetime metric $g_{\mu\nu}$. The repeated indices mean sum throughout the manuscript unless otherwise stated. The symmetric and trace-free part of a tensor $T^{i_{1}i_{2}i_{3}\cdots i_{n}}$ is denoted as $T^{<i_{1}i_{2}i_{3}\cdots i_{n}>}$. The time derivative of an object is represented by a dot over the corresponding variable.
\section{Relaxed EFE and waveform}\label{Sec.REFE}

%
To solve the EFE~\eqref{EFE} in the weak-field limit we use the DIRE approach~\cite{1975ApJ...197..717E} (see also~\cite{PhysRevD.54.4813,PhysRevD.62.124015}). First, we introduce the gothic metric $\mathfrak{g}^{\mu\nu}:=\sqrt{-g}g^{\mu\nu}$. Then, we define the tensor
\begin{equation}   H^{\alpha\mu\beta\nu}:=\mathfrak{g}^{\mu\nu}\mathfrak{g}^{\alpha\beta}-\mathfrak{g}^{\alpha\nu}\mathfrak{g}^{\beta\nu} \,,
\end{equation}
where the following identity holds:
\begin{eqnarray}\label{identity_H}
    \partial_{\mu}\partial_{\nu}H^{\mu\alpha\nu\beta}&=&(-g)\left( 2G^{\alpha\beta}+\frac{16\pi G}{c^4}t_{\mathrm{LL}}^{\alpha\beta} \right) \,, 
\end{eqnarray}
here $t_{\mathrm{LL}}^{\alpha\beta}$ is the Landau-Lifshitz energy-momentum tensor~\cite{Landau:1975pou}: 
\begin{widetext}
\begin{eqnarray}\label{Landau_Lifshitz_tensor}
\frac{16\pi G}{c^4}(-g)t_{LL}^{\alpha\beta}&=&g_{\lambda\nu}g^{\nu\rho}\partial_{\nu}\mathfrak{g}^{\alpha \lambda}\partial_{\rho}\mathfrak{g}^{\beta \mu}+\frac{1}{2}g_{\lambda\mu}g^{\alpha\beta}\partial_{\rho}\mathfrak{g}^{\lambda\nu}\partial_{\nu}\mathfrak{g}^{\rho\mu}-g_{\mu\nu}\left( g^{\lambda\alpha}\partial_{\rho}\mathfrak{g}^{\beta\nu}+g^{\lambda\beta}\partial_{\rho}\mathfrak{g}^{\alpha\nu} \right)\partial_{\lambda}\mathfrak{g}^{\rho\mu} \nonumber \\
&& + \frac{1}{8}\left( 2g^{\alpha\lambda}g^{\beta\mu}-g^{\alpha\beta}g^{\lambda\mu} \right)\left( 2g_{\nu\rho}g_{\sigma\tau}-g_{\rho\sigma}g_{\nu\tau} \right)\partial_{\lambda}\mathfrak{g}^{\nu\tau}\partial_{\mu}\mathfrak{g}^{\rho\sigma} \,,
\end{eqnarray}
\end{widetext}
and the Einstein tensor $G_{\mu\nu}$ is defined as:
%
%
\begin{equation}
G_{\mu\nu}:= R_{\mu\nu}-\frac{1}{2}g_{\mu\nu}R = \frac{8\pi G}{c^4}T_{\mu\nu}-\Lambda \mathfrak{g}^{-1/2}\mathfrak{g}_{\mu\nu} \,,
\end{equation}
with $\mathfrak{g}:=\mathrm{det}(\mathfrak{g}^{\mu\nu})=(-g)$. In order to study the field outside the source we expand the gothic metric around the Minkowski metric as follows:
\begin{equation}\label{y_decomposition}
    \mathfrak{g}^{\mu\nu}=\eta^{\mu\nu}+h^{\mu\nu} \,,
\end{equation}
where $h^{\mu\nu}$ stands as a potential. We select the harmonic gauge~\cite{Maggiore:2007ulw,poisson2014gravity}: $\partial_\mu \mathfrak{g}^{\mu\nu}=0$; therefore the relation \eqref{identity_H} becomes the wave equation:
\begin{equation}\label{REFE}
    \Box h^{\alpha\beta}=\frac{16\pi G}{c^4}\mu^{\alpha\beta} \,,
\end{equation}
where $\mu^{\alpha\beta}$ is the source of the system, which is given by:
%
%
\begin{equation}\label{mu_alpha_beta}
\mu^{\alpha\beta} = (-g)T^{\alpha\beta}+\frac{c^4}{16\pi G}\Lambda_{\mathrm{GR}}^{\alpha\beta} \,, 
\end{equation}
and in this approach the cosmological constant is taken as:
\begin{eqnarray}\label{source_LL_tensor_cosmological}
\Lambda_{\mathrm{GR}}^{\alpha\beta}&:=&\frac{16\pi G}{c^4}(-g)t_{LL}^{\alpha\beta}-2\Lambda \mathfrak{g}^{-1/2}\mathfrak{g}^{\alpha\beta}+ \partial_{\mu}h^{\alpha\mu}\partial_{\nu}h^{\beta\nu} \nonumber \\
    &-&h^{\mu\nu}\partial_{\mu}\partial_{\nu}h^{\alpha\beta} \,.
\end{eqnarray}
The previous expression \eqref{REFE} is known as the relaxed EFE.
%
\subsection{The tensor waveforms at 1PN}
In this subsection we compute the near-zone contribution wave form of $h^{ij}$. The near zone contribution of the faraway zone term is identified as $h_{\mathrm{N}}^{ij}$; which is given by~\cite{1975ApJ...197..717E,PhysRevD.54.4813}:
\begin{equation}\label{far_wave_form}
    h_{N}^{ij}(x)=\frac{2G}{Rc^4}\frac{d^2}{dt^2}\sum_{l=0}^{\infty}\hat{N}_{k_{1}}\cdots \hat{N}_{k_{l}} I_{EW}^{ijk_{{1}}\cdots k_{l}} \,,
\end{equation}
where $\hat{N}_{k_{l}}$ is the unit normal vector pointing from the source to the detector and $R$ is the distance between the source and the detector. All terms of $I_{\mathrm{EW}}^{ijk_{1}\cdots k_{l}}$ are known as the EW moments; and they are given explicitly as:
\begin{eqnarray}
   \label{EWij} I_{\mathrm{EW}}^{ij}&:=& \frac{1}{c^2}\int_{M}\mu^{00}x^{i}x^{j}d^3 x \,, \\
    \label{EWijk}I_{\mathrm{EW}}^{ijk}&:=&\frac{1}{c^3}\int_{M}\left( 2\mu^{0(i}x^{j)}x^{k}-\mu^{0k}x^i x^j \right)d^3 x \,, \\
   \hspace{-1.0cm} \label{EWijkl}I_{\mathrm{EW}}^{ijk_{1}\cdots k_{l}}&:=&\frac{2}{l!c^2}\frac{d^{l-2}}{d(ct)^{l-2}}\int_{M}\mu^{ij}x^{k_{1}}x^{k_{2}}\cdots x^{k_l} d^3 x \,.
\end{eqnarray}
To determine the GWs at 1PN order, we have to compute up to the fourth index of the EW moments, keeping in mind the transverse-traceless (TT) gauge~\cite{PhysRevD.54.4813} of the spatial tensor, that is:
\begin{equation}
    h^{ij}_{N}(x)=\frac{2G}{Rc^4}\frac{d^2}{dt^2}\left \lbrace  I^{ij}+\hat{N}_{k}I^{ijk}+\hat{N}_{k}\hat{N}_{l}I^{ijkl} \right\rbrace_{\mathrm{TT}} \,.
\end{equation}
Moreover, the $\mathrm{TT}$ operator acting on a tensor object $A^{ij}$ is such that~\cite{Maggiore:2007ulw, poisson2014gravity}:
\begin{eqnarray}
    \label{h_TT_projection}h^{ij}_{\mathrm{TT}}&=& \left( P^{ik}P^{jl}-\frac{1}{2}P^{ij}P^{kl} \right)A_{kl}, \\
    \label{P_operator}P^{ij}&:=&\delta^{ij}-\hat{N}^i \hat{N}^j \,,
\end{eqnarray}
where $P^{ij}$ is a operator projection and satisfying the properties $P^{ii}=2$, $P^{ij}P_{ij}=2$, and $P^{ij}P^{ik}=P^{jk}$.

In order to compute the EW moments, we need to know the explicit form of the source components $\mu^{\alpha\beta}$ to the $1$PN order. To do so, we have to obtain the terms $h^{00}$, $h^{0i}$, and $h^{ij}$ at the lowest order 
from the relaxed EFE, where the energy-momentum tensor has the following form:
\begin{equation}\label{EM_tensor_two_bodies}
    T^{\mu\nu}= \frac{1}{\sqrt{-g}}\sum_a m_a \frac{d\tau_a}{dt}\frac{dx_a^{\mu}}{d\tau_a}\frac{dx_a^{\nu}}{d\tau_a}\delta^3 (\Vec{x}-\Vec{x}_a(t)) \,,
\end{equation}
where $\delta^{3}(\vec{x}-\vec{x}_a(t))$ is the three dimensional Dirac delta, $m_a$ is the mass of the particle $a$, $\tau_a$ is the proper time of the particle $a$ and $t$ is the coordinate time. Therefore, from the relaxed EFE~\eqref{REFE}, the equations of motion become:
\begin{eqnarray}
    \label{eq_h_00}\nabla^2 h^{00}&=& \frac{16\pi G}{c^2}\sum_a m_a \delta^3(\Vec{x}-\Vec{x}_a(t))+2\Lambda \nonumber \\
    &+&O\left(\Lambda h, \frac{1}{c^4}\right) \,,\\
   \label{eq_h_0i} \nabla^2 h^{0i}&=& O\left(\Lambda h, \frac{1}{c^3}\right) \,, \\
   \label{eq_h_ij} \nabla^{2}h^{ij}&=& -2\Lambda \eta^{ij}+O\left(\Lambda h, \frac{1}{c^4}\right) \,.
\end{eqnarray}
Observe that in equation~\eqref{eq_h_00} we can globally factorize the coefficient $1/c^{2}$, therefore the second term, that is, the Newtonian order (lowest order), corresponds to the expression $c^2 \Lambda$. Thus, $\Lambda$ plays the role of a PN factor. 

It is worth to mention that in this approach the field $h_{\mu\nu}$ and $\Lambda$ are considered as perturbations, namely, we are considering a very small positive cosmological constant~\cite{PhysRevD.81.084002}. Hence, the equations of motion of all $h^{\alpha\beta}$ are taken at the lowest order and neglecting the terms $O(\Lambda h^{\mu\nu})$; and the harmonic gauge must be met, i.e., $\partial_{\mu}h^{\mu\nu}=0$. We point out that the lowest order of the temporal and spatial components are $O(c^{-3},\Lambda c^{-1})$ and $O(1,\Lambda)$, respectively. Therefore, the solutions of the $h^{\mu\nu}$ components read:
\begin{eqnarray}
   \label{h00} h^{00}&=&-\frac{4G}{c^2}\sum_a \frac{m_a}{r_a}+\frac{\Lambda}{3}|\Vec{x}|^2+O(\Lambda h,c^{-4}) \,, \\
   \label{h0i} h^{0i}&=& O(\Lambda h, c^{-3}) \,, \\
   \label{hij} h^{ij}&=&\delta^{ij}\left[ -\frac{1}{2}\Lambda \left( |\Vec{x}|^2-x_i^2 \right) \right]+O(\Lambda h, c^{-4}) \,,
\end{eqnarray}
with no sum in the term $x_i$ from $h^{ij}$~\eqref{hij}. Notice that the trace of the spatial component reads $h^{ii}=-\Lambda|\Vec{x}|^2$. In explicit matrix form the solution $h^{\mu\nu}$ is written as:
\begin{widetext}
\begin{equation}\label{matrix_form}
h^{\mu\nu}=
\begin{pmatrix}
-\frac{4G}{c^2}\sum_a \frac{m_a}{r_a}+\frac{\Lambda}{3}|\Vec{x}|^2 & 0 & 0&0\\
0 & -\frac{1}{2}\Lambda(y^2+z^2) & 0&0\\
0 & 0 & -\frac{1}{2}\Lambda (x^2+z^2)&0 \\
0&0&0& -\frac{1}{2}\Lambda (x^2+y^2)
\end{pmatrix} \,,
\end{equation}
\end{widetext}
where $x$, $y$ and $z$ are the Cartesian coordinates. The condition $\partial_{\mu}h^{\mu0}=0$ holds as long as we add $h^{0i}$ at order $O(\Lambda h,c^{-3})$ and $h^{00}$ (given in equation \eqref{h00}); whilst for the case $\partial_{\mu}h^{\mu i}=0$ we only need to consider $h^{ij}$, which is given in equation \eqref{hij}. Moreover, we highlight that the solution of $h^{\mu\nu}$ has a cylindrical symmetry around the corresponding principal axis. This symmetry is a remnant  of the rotational symmetry, making that the shattering of this symmetry be an artifact of the harmonic gauge condition~\cite{PhysRevD.81.084002}.

From the definition of the gothic metric and its expansion~\eqref{y_decomposition} we can obtain the components of the metric at the same order of the perturbation $h^{\mu\nu}$. Accordingly, we have:
\begin{equation}
(-g)=\mathrm{det}(\mathfrak{g}^{\mu\nu}) = \mathrm{det}(\eta^{\mu\nu}+h^{\mu\nu}) = 1+h+O(h^2) \,,     
\end{equation}
%
%
with $h:=h^{\mu\nu}\eta_{\mu\nu}$ and $\mathfrak{g}_{\mu\nu}=\eta_{\mu\nu}-h_{\mu\nu}+O(h^2)$. Thus, the components of the metric at lowest order read:
\begin{eqnarray}
&& \hspace{-1cm} \label{g_00_h}  g_{00} = \sqrt{-g}\mathfrak{g}_{00} \nonumber \\
&& \hspace{-0.5cm} = -(1+\frac{1}{2}h^{00}+\frac{1}{2}h^{ii})+O(h^2, \Lambda h) \,, \\
&& \hspace{-1cm} \label{g_0i_} g_{0i} = \sqrt{-g}\mathfrak{g}_{0i} \nonumber \\
&& \hspace{-0.5cm} = O(\Lambda h, c^{-3}) \,, \\
&& \hspace{-1cm} \label{g_ij_h}  g_{ij} = \sqrt{-g}\mathfrak{g}_{ij} \nonumber\\
&&  \hspace{-0.5cm} = \eta_{ij}-h_{ij}-\frac{1}{2}\eta_{ij}h^{00}+\frac{1}{2}h^{kk}\eta_{ij}+O(h^2, \Lambda h) \,,
\end{eqnarray}
%
%
here we have used $h=-h^{00}+h^{ii}$. When substituting the components of the metric into those of the source~\eqref{mu_alpha_beta}, it yields
\begin{widetext}
\begin{eqnarray}
 \label{mu_00}   \mu^{00}&=& c^2 \sum_a m_a\left( 1-\frac{3}{4}h^{00} +\frac{1}{4}h^{ii}+\frac{v_a^2}{2c^2} \right)\delta^3(\Vec{x}-\vec{x}_a(t)) +\frac{c^4}{16\pi G}\left[ -\frac{7}{8}\partial_i h^{00}\partial_i h^{00}+2\Lambda  \right]+O(\Lambda h, c^{-2}) \,, \\
  \label{mu_0i}  \mu^{0i}&=& c\sum_a m_a v_a^i \delta^3(\Vec{x}-\Vec{x}_a(t))+O(\Lambda h, c^{-2}) \,, \\
   \label{mu_ij} \mu^{ij}&=& \sum_a m_a v_a^i v_a^j \delta^3(\Vec{x}-\Vec{x}_a(t))+\frac{c^4}{16\pi G}\left[ \frac{1}{4}\partial^i h^{00}\partial^j h^{00}-\frac{1}{8}\delta^{ij}\partial_k h^{00}\partial_k h^{00} \right]-2\Lambda \delta^{ij}+O(\Lambda h, c^{-2}) \,,
\end{eqnarray}
\end{widetext}
with $v_a$ as the velocity of the particle $a$. We stress that the cosmological constant only appears in the later terms of $\mu^{00}$ and $\mu^{ij}$. In the case of $\Lambda=0$, one recovers the usual gravitational case at $1$PN order~\cite{PhysRevD.62.124015}.
\section{Evaluation of the Epstein-Wagoner moments}
From the expressions \eqref{EWij}, \eqref{EWijk}, and \eqref{EWijkl}, one can observe that they are volume integral over a sphere of radius $\mathcal{R}$, which is the boundary of the near region, evaluated at the retarded time $\tau=t-R/c$. In this section we present some steps, in detail, in order to obtain the EW moments, particularly we focus in the terms that contain $\Lambda$, since the integrals that do not have this parameter are already evaluated in \cite{PhysRevD.54.4813}. Also in various occasions we integrate by parts and we use the identity~\cite{PhysRevD.54.4813, poisson2014gravity}:
\begin{equation}
    \int_{M}\partial_k F^{ij \cdots k}d^3x=\int_{\partial M}F^{ij\cdots k}|_{\mathcal{R}}\hat{N}_{k}\mathcal{R}^2 d \Omega \,,
\end{equation}
with $\partial M$ as the boundary of the $3$ dimensional manifold $M$ at the near region, and $\mathcal{R}^2 \hat{n}^k d\Omega^2=dS^k$ is the surface element at this border. Moreover, we are only interested in finding the tensorial terms that survive after applying the $\mathrm{TT}$ projector. Thus, the subsequent identities follow from the projection operator $P^{i}{}_{j}$, given in equation \eqref{h_TT_projection} and \eqref{P_operator},
\begin{eqnarray}
(\delta^{ij})_{\mathrm{TT}} &=& 0 \,, \\
(\hat{N}^i F^{j})_{\mathrm{TT}} &=& 0 \,,
\end{eqnarray}
where the indices $i$ and $j$ apply to the final components of the waveform (not the integrands), and $F$ denotes a general term. All these results and procedures are explained in detail in \cite{PhysRevD.54.4813, poisson2014gravity}.
\subsection{Two indices moment $I_{\mathrm{EW}}^{ij}$}
The computation of the moment $I_{\mathrm{EW}}^{ij}$ begins using the following useful identity:
\begin{eqnarray}
\hspace{-0.5cm} \label{id_1} \partial_k h^{00}\partial_k h^{00}x^ix^j&=&\partial_k\left( h^{00}\partial_k h^{00}x^ix^j \right) \nonumber\\
&& \hspace{-1.2cm} - h^{00}\left( \nabla^{2}h^{00}x^ix^j+\partial^ih^{00}+\partial^jh^{00}x^i \right) \,,
\end{eqnarray}
and after some algebraic manipulation we obtain:
\begin{eqnarray}
\hspace{-0.5cm} \label{id_2} -h^{00}\partial^ih^{00}x^j-h^{00}\partial^jh^{00}x^i&=&-\frac{1}{2}\partial^i[(h^{00})^2x^j] \nonumber\\
&& \hspace{-1.2cm} -\frac{1}{2}\partial^j[(h^{00})^2x^i]+(h^{00})\delta^{ij} \,.
\end{eqnarray}
Substituting equation \eqref{mu_00} into \eqref{EWij}, then using equation \eqref{id_1} and \eqref{id_2}, neglecting the boundary terms and considering the $\mathrm{TT}$ components of the far zone tensor perturbation, it yields:
\begin{eqnarray}
\label{EW_1ij}I_{\mathrm{EW}}^{ij}&\overset{\mathrm{TT}}{=}& \frac{1}{c^2}\int_{M}\bigg \lbrace c^2 \sum_a m_a\left( 1-\frac{3}{4}h^{00} +\frac{1}{4}h^{ii} \right. \nonumber\\
&& +\left. \frac{v_a^2}{2c^2} \right)\delta^3(\Vec{x}-\vec{x}_a(t)) \bigg \rbrace x_a^{<i}x_a^{j>} d^3x \nonumber\\
&& +\frac{c^4}{c^2 16\pi G}\frac{7}{8}\int_{M} h^{00}\nabla^2 h^{00}x^{<i}x^{j>} d^3x \,.
\end{eqnarray}
Next, we use the results \eqref{eq_h_00} and \eqref{h00}, so the last term becomes:
\begin{widetext}
\begin{eqnarray}
    \frac{c^2}{16\pi G} \frac{7}{8}\int_{M}h^{00}\nabla^2 h^{00}x^{<i}x^{j>}d^3x &\overset{\mathrm{TT}}{=}& \frac{c^2}{16\pi G}\frac{7}{8}\int_{M}\bigg( -\frac{4G}{c^2}\sum_a \frac{m_a}{r_a} +\frac{\Lambda}{3}|\vec{x}|^2 \bigg)\nabla^2 h^{00} x^{<i}x^{j>}d^3x \nonumber\\
    &\overset{\mathrm{TT}}{=}&\frac{c^2}{16\pi G}\frac{7}{8}\int_M \bigg( -\frac{4G}{c^2}\sum_a \frac{m_a}{r_a} \nabla^2 h^{00}x^{<i}x^{j>}\bigg) d^3x \nonumber\\
    &\overset{\mathrm{TT}}{=}&-\frac{7G}{2c^2}\sum_a \sum_{b\neq a}\frac{m_b}{r_{ab}}x_a^{<i}x_a^{j>} \,,
\end{eqnarray}
\end{widetext}
where we have neglected all the terms with $\Lambda h$. Finally, plugging in \eqref{h00} and \eqref{hij} into the first term of \eqref{EW_1ij} and putting together the previous result, the two indices EW moment becomes:
\begin{eqnarray}
    \label{two_index_moment} I_{\mathrm{EW}}^{ij}&\overset{\mathrm{TT}}{=}&\sum_a m_a \left[ 1-\frac{G}{2c^2}\sum_{b\neq a}\frac{m_b}{r_{ab}}-\frac{\Lambda}{2}|\Vec{x}_a|^2+\frac{v_a^2}{2c^2} \right]x_a^{i}x_a^{j} \nonumber\\
    && + O(\Lambda h, c^{-4}) \,.  
\end{eqnarray}
\subsection{Three indices moment $I_{\mathrm{EW}}^{ijk}$}
In this case since there is no cosmological constant contribution, the computation of the integral is direct, yielding:
\begin{eqnarray}
    \label{three_index_moment}I_{\mathrm{EW}}^{ijk}&\overset{\mathrm{TT}}{=}&\frac{1}{c}\sum_a m_a \left[ 2v_a^{(i}x_a^{j)}x_a^{k}-v_a^{k}x_a^i x_a^j  \right] \nonumber\\
    && + O(\Lambda h, c^{-3}) \,. 
\end{eqnarray}
\subsection{Four indices moment $I_{\mathrm{EW}}^{ijkl}$}
Plugging in the spatial component of the source \eqref{mu_ij} into the corresponding four indices integral \eqref{EWijkl} and considering the $\mathrm{TT}$ gauge, we obtain:
\begin{eqnarray}
    I_{\mathrm{EW}}^{ijkl}&\overset{\mathrm{TT}}{=}&\frac{1}{c^2}\sum_a m_a v_a^i v_a^j x_a^k x_a^l \nonumber\\
    && + \frac{c^2}{64\pi G}\int_{M}\partial^i h^{00}\partial^{j}h^{00}x^{k}x^{l}d^{3}x \,.
\end{eqnarray}
Next, we integrate by parts the last term, neglecting all the boundary terms, applying the $\mathrm{TT}$ gauge and using \eqref{h00}, yielding:
\begin{eqnarray}
&&\hspace{-1cm} \int_M \partial^i h^{00}\partial^j h^{00}x^kx^l d^3x \overset{\mathrm{TT}}{=}
-\int_M h^{00}\partial^i \partial^j h^{00}x^{k}x^{l}d^3 x \nonumber\\
&&\hspace{-1cm} \quad\quad\quad\quad\quad\quad\overset{\mathrm{TT}}{=}\frac{4G}{c^2}\int_M \sum_a \frac{m_a}{r_a}\partial^i \partial^j h^{00}x^k x^l d^3 x \,, 
\end{eqnarray}
here in the last term we have neglected expressions with $\Lambda h$; and also the $\mathrm{TT}$ gauge projection was not used. On the other hand, using again the result \eqref{h00} one obtains:
\begin{eqnarray}
    \partial_i\partial_jh^{00}&=&\frac{4G}{c^2}\sum_a m_a \bigg[ \frac{1}{|\vec{x}-\vec{x}_a|^3}\delta_{ij} \nonumber\\
    &&\hspace{-1.0cm} -\frac{3}{|\vec{x}-\vec{x}_a|^5}(\vec{x}-\vec{x}_{a})_i(\vec{x}-\vec{x}_a)_j \bigg] +\frac{2}{3}\Lambda \delta_{ij} \,,
\end{eqnarray}
with $\vec{x}\neq \vec{x}_a$. After applying the $\mathrm{TT}$ gauge the cosmological constant term vanishes. Therefore, it turns out that the four indices EW moment has no $\Lambda$; thus from \cite{PhysRevD.54.4813}, the integral reads:
\begin{eqnarray}
\label{four_index_moment}I_{\mathrm{EW}}^{ijkl}&\overset{\mathrm{TT}}{=}& \frac{1}{c^2}\sum_a m_a v_a^i v_a^j x_a^k x_a^l \nonumber\\
    &+&\sum_a \sum_{b\neq a}\bigg[ \frac{G m_a m_b}{12 rc^2}x^i x^j \left( \frac{x^k x^l}{|\Vec{x}|^3}-\delta^{kl}-6\frac{x_a^k x_a^l}{|\Vec{x}|^2} \right) \bigg] \nonumber\\
    &+&O(\Lambda, hc^{-4}) \,. 
\end{eqnarray}
%
%
We remark that in the three and four indices moments there are no contribution of the cosmological constant under the TT projection; however, $\Lambda$ will have an impact on higher order approximations. 
\subsection{Center of mass at $1$PN order with $\Lambda$}
To express the waveform in terms of the relative variables we move to the center of mass frame $X_{\mathrm{CM}}^i=0$, that is:
\begin{eqnarray}
\hspace{-1.5cm} X^{i}_{\mathrm{CM}}&:=& \frac{1}{m}\int_{M} \mu^{00}x^{i}d^3 \Vec{x} \nonumber \\
    &=& \frac{1}{m}\sum_a m_a x_a^i-\frac{G}{2c^2 m}\sum_a m_a \sum_{b\neq a}\frac{m_b}{r_{ab}}x_a^i \nonumber \\
    && +\frac{1}{2c^2 m}\sum_a m_a v_a^2 x_a^i-\frac{\Lambda}{2m}\sum_a m_a |\Vec{x}_a|^2 x_a^i \nonumber \\
    && + O(\Lambda h, c^{-3}) \,,
\end{eqnarray}
where we have used equations \eqref{h00} and \eqref{mu_00}, and the spatial trace of equation \eqref{hij}. On the other hand, considering only two particles in interaction, we find at $1$PN order that the coordinates of each body in the center of mass frame is given by:
\begin{eqnarray}
    \label{y_1}\Vec{r}_1&=& \frac{\mu}{m_1}\Vec{r}+\frac{\mu \Delta m}{2m^2 c^2}\left( v^2 -\frac{Gm}{r}-\frac{\Lambda c^2 r^2}{2} \right)\Vec{r} \nonumber \\
    && + O(c^{-2}\Lambda, \Lambda^2, c^{-3}) \,, \\
    \label{y_2}\Vec{r}_2&=& -\frac{\mu}{m_2} \Vec{r}+\frac{\mu\Delta m}{2m^2 c^2}\left( v^2 -\frac{Gm}{r}-\frac{\Lambda c^2 r^2}{2} \right)\Vec{r} \nonumber \\
    && + O(\Lambda c^{-2}, \Lambda^2, c^{-3}) \,,
\end{eqnarray}
with $\Vec{r}=\Vec{r}_1-\Vec{r}_2$ as the relative position, $r=|\Vec{r}|$, $\Vec{v}=\Vec{v}_1-\Vec{v}_2$ as the relative velocity, $v=|\Vec{v}|$, $\mu=m_{1}m_{2}/m$ as the reduced mass of the binary system, $m=m_{1}+m_{2}$, and $\Delta m:=m_1-m_2$. 

Computing the time derivative of the positions given by equations \eqref{y_1} and \eqref{y_2} leads to the velocities of each particle:
\begin{eqnarray}
    \label{v_1}\Vec{v}_1&=& \frac{\mu}{m_1}\Vec{v}+\frac{\mu \Delta m}{2m^2 c^2}\left[ \left( v^2-\frac{Gm}{r}-\frac{\Lambda c^2 r^2}{2} \right)\Vec{v} \right. \nonumber \\
    &-& \left. \left( \frac{Gm}{r^2}+\frac{\Lambda c^2 r}{2} \right) \dot{r}\Vec{r} \right]+O(\Lambda c^{-2}, \Lambda^2, c^{-4}) \,, \\
    \label{v_2}\Vec{v}_2&=& -\frac{\mu}{m_2}\Vec{v}+\frac{\mu \Delta m}{2m^2 c^2}\left[ \left( v^2-\frac{Gm}{r}-\frac{\Lambda c^2 r^2}{2} \right)\Vec{v}\right. \nonumber\\
    &-&\left. \left( \frac{Gm}{r^2}+\frac{\Lambda c^2 r}{2} \right) \dot{r}\Vec{r} \right]+O(\Lambda c^{-2}, \Lambda^2, c^{-4}) \,.
\end{eqnarray}
Now we substitute the positions and velocities \eqref{y_1}, \eqref{y_2}, \eqref{v_1}, and \eqref{v_2} into the EW moments \eqref{two_index_moment}, \eqref{three_index_moment}, and \eqref{four_index_moment}; finally we plug in the whole thing in equation \eqref{far_wave_form}, obtaining the waveform of a two body compact system in a general motion:
\begin{eqnarray}
  h^{ij}_{\mathrm{N}, \mathrm{TT}}&=& \frac{2G \mu}{Rc^4}\frac{d^2}{dt^2}\bigg \lbrace \bigg [ 1+\frac{1}{2c^2}(1-3\nu)(v^2-\Lambda c^2 r^2) \nonumber\\
    && \hspace{-1.0cm} - \frac{Gm}{3rc^2}(2-9\nu)\bigg] r^i r^j -\frac{\Delta m}{mc^2}\bigg( 2v^{(i}r^{j)}(\hat{N}\cdot \Vec{r}) \nonumber\\
    && \hspace{-1.0cm} - (\hat{N}\cdot \Vec{v})r^i r^j \bigg) \nonumber\\
    && \hspace{-1.0cm} + \frac{1}{c^2}(1-3\nu)(\hat{N}\cdot \Vec{r})^2 \bigg( v^i v^j -\frac{Gm}{3r^3}r^i r^j \bigg) \bigg \rbrace _{\mathrm{TT}} \,,
\end{eqnarray}
with $\nu:=\mu/m=m_{1}m_{2}/m^2$ as the symmetric mass ratio. Then we take the time derivative and we use the relative $1$PN acceleration \eqref{EOM} (which was computed in appendix \ref{two_lagrangian}) where required, we arrive at the final form of the near zone waveform:
\begin{eqnarray}\label{wave_form_complete}
 h^{ij}_{\mathrm{N,TT}}(t,x)
 &=& \frac{2G\mu}{c^4R}\bigg \lbrace \Tilde{Q}^{ij}+\frac{1}{c}P^{1/2}\Tilde{Q}^{ij}+\frac{1}{c^2}P\Tilde{Q}^{ij} \nonumber\\
 && + O(c^{-3},c^{-1}\Lambda, \Lambda^2) \bigg\rbrace \,,
\end{eqnarray}
with
\begin{widetext}
\begin{eqnarray}
\Tilde{Q}^{ij}&=& 2\left( v^iv^j-\frac{Gm}{r^3}r^ir^j \right)+\frac{\Lambda}{3}c^2 r^i r^j \,, \\
P^{1/2}\Tilde{Q}^{ij}&=&\Delta m\bigg [ 3\frac{Gm}{r^3}(\hat{N}\cdot \Vec{r})\left( 2v^{(i}r^{j)}-\frac{\dot{r}}{r}r^ir^j \right)+(\Vec{v}\cdot \hat{N})\left( -2v^iv^j+\frac{Gm}{r^3}r^ir^j \right)-2\Lambda c^2 (\hat{N}\cdot \Vec{r})v^{(i}r^{j)} \nonumber \\
&& - \frac{\Lambda}{3}c^2(\hat{N}\cdot  \Vec{v})r^i r^j \bigg]  \,, \\
P\Tilde{Q}^{ij}&=&  \frac{1}{3}\left[ 3(1-3\nu)v^2-2(2-3\nu)\frac{Gm}{r} \right]v^iv^j+\frac{4}{3}(5+3\nu)\frac{Gm}{r^2}\dot{r}v^{(i} r^{j)} \nonumber \\
&& +\frac{1}{3}\frac{Gm}{r^3}\left[ -(10+3\nu)v^2+3(1-3\nu)\dot{r}^2+29\frac{Gm}{r} \right]r^i r^j+\frac{2}{3}(1-3\nu)(\Vec{v}\cdot \hat{N})^2 \left( 3v^i v^j-\frac{Gm}{r^3}r^i r^j \right) \nonumber \\
&& + \frac{4}{3}(1-3\nu)(\Vec{v}\cdot \hat{N})(\Vec{r}\cdot\hat{N})\frac{Gm}{r^3}\left[ -8v^{(i}r^{j)}+3\frac{\dot{r}}{r}r^i r^j \right] \nonumber \\
&& + \frac{1}{3}(1-3\nu)(\Vec{r}\cdot \hat{N})^2\frac{Gm}{r^3} \bigg[ -14 v^i v^j+30\frac{\dot{r}}{r}v^{(i}r^{j)}+\left( 3\frac{v^2}{r^2}-15\frac{\dot{r}^2}{r^2}+7\frac{Gm}{r^3} \right)r^i r^j \bigg]  \nonumber \\
&& - \frac{17\Lambda c^2}{9}(1+3\nu)\frac{Gm}{r}r^i r^j-\Lambda c^2 \left[ 2\left( \frac{2}{3}-\nu \right)v^2+(1-3\nu)\left( \frac{Gm}{r^3}(r_i)^2+(v_i)^2 \right) \right]r^i r^j \nonumber \\
&& + \Lambda c^2 \left[ 2(1-3\nu)r_i \dot{r}_i-(6-14\nu)r\dot{r} \right]v^{(i}r^{j)}-\Lambda c^2 (1-3\nu)r^2 v^i v^j+\frac{4}{3}\Lambda c^2 (1-3\nu)(\hat{N}\cdot \Vec{r})^2 v^i v^j \nonumber \\
&& - \frac{13}{9}\Lambda c^2(1-3\nu)\frac{Gm}{r^3}(\hat{N}\cdot \Vec{r})^2r^i r^j+\frac{8}{3}\Lambda c^2(1-3\nu)(\hat{N}\cdot \Vec{v})(\hat{N}\cdot \Vec{r})r^{(i}v^{j)} \,,
\end{eqnarray}
\end{widetext}
where the repeated indices do not indicate sum. For instance for $i=j=1$, we have:
\begin{eqnarray}
    \bigg( \frac{Gm}{r^3}(r_{i})^2+(v_i)^2 \bigg) r^ir^j &=&  \bigg( \frac{Gm}{r^3}(r_1)^2+(v_1)^2 \bigg)r^1r^1 \nonumber\\
    &=&\bigg( \frac{Gm}{r^3}x^2+(v_x)^2 \bigg) x^2 \,,
\end{eqnarray}
with $x$ as the Cartesian coordinate of the relative position $\vec{r}$ and $v_x$ as their respective velocity component (see Appendix~\ref{two_lagrangian}). Notice that omitting the cosmological constant the wave expression becomes the case of the gravitational radiation at $1$PN order~\cite{poisson2014gravity}. 
\section{Circular orbit}\label{Sec_Circ_orbit}
In this section we study the interaction of the two compact body system given the particular case of a circular orbit, which is the most simple case to analyze. Here we have to consider that $\Dot{r}=\ddot{r}=0$, as well as we denote $\Dot{\phi}:=\omega$ as the frequency of the system. We take into account some results obtained in appendix \ref{two_lagrangian}. Hence, the equation of motion \eqref{r_eq} is:
\begin{eqnarray}\label{frequency_two_bodies}
    \omega^2&=& \frac{Gm}{r^3}-\frac{\Lambda}{3}c^2-\frac{Gm}{c^2 r}\bigg[ \frac{Gm}{r^3}(3-\nu)-\frac{c^2\Lambda}{6}(10-3\nu) \bigg]\nonumber\\
    && + O(c^{-4},\Lambda c^{-2}, \Lambda^2 ) \,.
\end{eqnarray}
Additionally, we know that the velocity is given by
\begin{eqnarray}\label{vel_cir_orbit}
    v^2&=& (r\omega)^2 \nonumber \\
    &=& \frac{Gm}{r}-\frac{\Lambda}{3}c^2 r^2 \nonumber\\
    &-&\frac{Gm}{c^2 r}\bigg[ \frac{Gm}{r}(3-\nu)-\frac{c^2\Lambda}{6}r^2(10-3\nu) \bigg] \nonumber\\
    &+&O(c^{-4}, \Lambda c^{-2}, \Lambda^2) \,.
\end{eqnarray}
We remark that $\Lambda$ appears explicitly in each corresponding term at $1$PN order of \eqref{y_1} and \eqref{y_2}. For a circular motion, this term does not vanishes since the velocity of a compact binary system with $\Lambda$ given in \eqref{vel_cir_orbit} at Newtonian order does not coincide with the terms in brackets of equations \eqref{y_1} and \eqref{y_2}. In contrast with the case absence of $\Lambda$, there is no contribution at $1$PN order for a circular motion~\cite{PhysRevD.51.5360}. The energy of the system is (see \eqref{energy}):
\begin{eqnarray}\label{energy_components_velocity}
    E&=& mc^2 -\frac{G \mu m}{2r}\bigg[1-\frac{1}{4}(7-\nu)\frac{Gm}{c^2 r} \bigg]\nonumber\\
    &-&\frac{1}{3}\mu\Lambda c^2 r^2+\frac{\Lambda}{6}\left( \frac{13}{2}+5\nu \right) \nonumber \\
    &-&\frac{11}{2}\Lambda \mu (1-3\nu)(x^2 v_x^2+y^2v_y^2+z^2v_z^2) \,.
\end{eqnarray}
Recalling the orbital plane coordinates:
\begin{eqnarray}
    x&=&r\mathrm{cos}\phi, \\
    y&=& r\mathrm{sin}\phi, \\
    z&=&1, 
\end{eqnarray}
then if $\dot{r}=0$, this leads to:
\begin{eqnarray}
    v_x&=& \dot{x}= -r\omega \mathrm{sin}\phi \,, \\
    v_y&=& \dot{y}= r\omega \mathrm{cos}\phi \,, \\
    v_z&=&\dot{z}=0 \,,
\end{eqnarray}
now we can obtain the following identity:
\begin{eqnarray}\label{identity_velocities}
\hspace{-1cm} x^2 v_x^2+y^2v_y^2+z^2v_z^2&=& 2r^4 \omega^2 \mathrm{cos}^2\phi \mathrm{sin}\phi \nonumber \\
    && \hspace{-1.0cm} = \frac{1}{2}r^2 v^2\mathrm{sin}^2(2\phi) \nonumber \\
    && \hspace{-1.0cm} = \frac{1}{2}Gmr \mathrm{sin}^2(2\phi) + O(c^{-2}, \Lambda c^2) \,, 
\end{eqnarray}
and here we have used \eqref{vel_cir_orbit}. Consequently, we substitute \eqref{identity_velocities} into the energy \eqref{energy_components_velocity}, resulting in:
\begin{eqnarray}\label{energy_two_body}
    E&=& mc^2-\frac{G\mu m}{2r}\bigg[ 1-\frac{1}{4}(7-\nu)\frac{Gm}{c^2 r} \bigg]-\frac{\Lambda}{3}\mu c^2 r^2 \nonumber \\
    && + \frac{\Lambda}{6}\mu Gmr(\frac{13}{2}+5\nu) \nonumber\\
    && - \frac{11}{4}\Lambda \mu Gmr(1-3\nu)\mathrm{sin}^2(2\phi) \nonumber \\
    && + O(c^{-4}, \Lambda c^{-2}, \Lambda^2) \,.
\end{eqnarray}
\subsection{Energy loss rate}
The flux of energy (see appendix~\ref{Radiated_power_formula}) that comes from the tensor wave is:
\begin{equation}\label{flux_energy_h_ij}
    P=\frac{c^3 R^2}{32\pi G}\int \dot{h}^{ij}_{\mathrm{TT}}\dot{h}_{ij}^{\mathrm{TT}}d^3 x \,,
\end{equation}
with $R$ as the distance from the source to the detector. In order to compute $P$, we can proceed in two different ways, in which we consider the particular case of a circular  orbit of a two body compact system, i.e., $\dot{r}=0$. A first approach is to differentiate $h^{ij}$ (from \eqref{wave_form_complete}) with respect time, where the $1$PN equation of motion \eqref{EOM} can be utilized, and we substitute this outcome into~\eqref{flux_energy_h_ij}. The other method is taking the appropriate time derivatives of the STF moments \eqref{I_circ_ij}, \eqref{I_circ_ijk}, and \eqref{J_circ_ij} (obtained in appendix \ref{STF}) and we plugin in them into~\eqref{radiated_power_STF}. Thus, the rate of loss of energy of such system is:   
%
%
\begin{eqnarray}\label{radiated_power_1}
    P&=&-\frac{G}{c^5}\frac{32}{5}(\nu m)^2\bigg[ \frac{G^3 m^3}{r^5}-\frac{G^2 m^2}{r^2}\Lambda c^2 \nonumber\\
    && - \frac{G^4 m^4}{c^2 r^6}\left( \frac{2927}{336}+\frac{5}{4}\nu \right) +\frac{G^3 m^3}{r^3}\Lambda \left( \frac{2423}{252}+\frac{31}{6}\nu \right) \nonumber\\
    && + O(c^{-4},c^{-2}\Lambda, \Lambda^2) \bigg] \,.  
\end{eqnarray}
\subsection{Post-Newtonian parameters}
For future purposes we introduce some PN parameters. The first one is defined as $\gamma:=\frac{Gm}{c^2r}$. Hence, the frequency of a circular orbit~\eqref{frequency_two_bodies} takes the following form:
\begin{equation}\label{freq_PN_gamma}
    \omega^2=\frac{Gm}{r^3}\bigg[ 1-\left( 3-\nu \right)\gamma \bigg]-\frac{\Lambda c^2}{3}\bigg[ 1-\frac{\gamma}{2}\left( 10-3\nu \right) \bigg] \,.
\end{equation}
Then, the relative distance can be expressed as:
\begin{eqnarray}
    r&=&\left( \frac{Gm}{\omega^2} \right)^{1/3}\left[ 1-\left( 3-\nu \right)\gamma \right]^{1/3}\times \nonumber\\
    &&\bigg\lbrace 1+\frac{\Lambda c^2}{3\omega^2}\left[ 1-\frac{\gamma}{2}\left( 10-3\nu \right) \right]  \bigg \rbrace^{-1/3} \,.
\end{eqnarray}
Therefore, the first PN parameter becomes: 
\begin{widetext}
    \begin{eqnarray}\label{gamma_in_terms_x}
    \gamma&=& \frac{Gm}{c^2 r} \nonumber\\
    &=& \left( \frac{\omega Gm}{c^3} \right)^{2/3}\left[ 1-(3-\nu)\gamma \right]^{-1/3}\bigg\lbrace 1+\frac{\Lambda c^2}{3\omega^2}\left[ 1-\frac{\gamma}{2}(10-3\nu) \right] \bigg\rbrace^{1/3} \nonumber \\
     &=& x^2 \bigg[ 1+\frac{1}{3}(3-\nu)x^2+\frac{\Lambda}{9}\frac{G^2m^2}{c^4 x^6}-\frac{1}{54}\frac{\Lambda G^2 m^2}{c^4 x^4}(6-\nu) +O(x^4,\Lambda c^{-4}x^{-8}, \Lambda^2) \bigg] \,,
    \end{eqnarray}
\end{widetext}
where the second PN parameter $x:=\left( \frac{\omega Gm}{c^3} \right)^{1/3}$, the inverse squared frequency given by:
\begin{equation}
    \omega^{-2}= \frac{r^3}{Gm}\left[ 1+(3-\nu)\gamma +O(\gamma^2, \Lambda c^2, \Lambda) \right] \,,
\end{equation}
which is obtained from \eqref{freq_PN_gamma} using Taylor series expansion, and the relation of both PN parameters $\gamma\simeq x^2$ were introduced. On the other hand, we introduce the PN parameter $\gamma$ from~\eqref{gamma_in_terms_x} into the radiated power~\eqref{radiated_power_1} and the energy~\eqref{energy_two_body}, yielding:
%
%
%
\begin{widetext}
\begin{eqnarray}
 \label{power_in_term_x}   P &=& -\frac{32}{5}\frac{c^5}{G}\nu^2 \gamma^5 \bigg[ 1-\gamma \left( \frac{2927}{336}+\frac{5}{4}\nu \right)-\frac{\Lambda G^2 m^2}{c^4 \gamma^3}+\frac{\Lambda G^2 m^2}{c^4 \gamma^2}\left( \frac{2423}{252}+\frac{31}{6}\nu \right) +O(\gamma^2, \Lambda c^{-4}\gamma^{-1}, \Lambda^2) \bigg] \nonumber\\
    &=&-\frac{32}{5}\frac{c^5}{G}\nu^2 x^{10}\bigg[ 1-\left( \frac{1247}{336}+\frac{35}{12}\nu \right)x^2-\frac{\Lambda G^2 m^2}{c^4 x^6}-\frac{\Lambda G^2 m^2}{432 c^4 x^4}(97-2692\nu) +O(x^4, \Lambda c^{-4}x^2, \Lambda^2) \bigg] \,, \\
    \label{energy_in_terms_x} E&=& mc^2 -\frac{\mu}{2}c^2\gamma \bigg[ 1-\frac{1}{4}(7-\nu)\gamma \bigg]-\frac{1}{3}\mu \Lambda \frac{G^2 m^2}{c^2 \gamma^2}+\frac{1}{6}\Lambda \mu \frac{G^2 m^2}{c^2\gamma}(\frac{13}{2}+5\nu)-\frac{11}{4}\Lambda \mu \frac{G^2m^2}{c^2\gamma}(1-3\nu)\mathrm{sin}^2(2\phi_{0\mathrm{PN}}) \nonumber \\
    &=& mc^2 -\frac{\mu}{2}c^2 x^2 \bigg[ 1-\left( \frac{3}{4}+\frac{1}{12}\nu \right)x^2+\frac{7}{9}\Lambda \frac{G^2 m^2}{c^4 x^6}-\frac{\Lambda G^2 m^2}{54 c^4 x^4}\left( 103+95\nu \right) \nonumber \\
    && + \frac{11}{2}\frac{\Lambda G^2 m^2}{c^4 x^4}(1-3\nu)\mathrm{sin}^2(2\phi_{0\mathrm{PN}}) +O(x^4, \Lambda c^{-4}x^2, \Lambda^2)
 \bigg] \,.
\end{eqnarray}
\end{widetext}
Here we have to remark that the energy depends explicitly of the phase of the circular orbit $\phi$ (inside the sine function), and the order of the complete calculation should be at $1$PN order; however, this information is not available at this stage, so in this approximation we introduce the Newtonian phase $\phi_{0\mathrm{PN}}$ that we already know at this point of the analysis. This occurrence of phase is a direct consequence of the fact that the spatial components of the waveform solution do not have rotational symmetry due to harmonic gauge artifact. Thus, the Newtonian phase in terms of the PN parameter $x$ is given by:
\begin{equation} \label{Newtonian_phase}
\phi_{0\mathrm{PN}} = -\frac{1}{32\nu}x^{-5}\bigg[ 1-\frac{25}{99}\frac{\Lambda G^2 m^2}{c^4}x^{-6}\bigg] \,.
\end{equation}
%
%
%
\subsection{Energy loss rate of a circular motion of a two body system}
Since the system is expected to loss energy, hence releasing GWs, this configuration becomes a binary quasicircular scenario. Then, to obtain the phase of the GW $\phi$ we must use the balance equation, namely:
\begin{equation}\label{balance_equation}
    \frac{dE}{dt}=-P \,.
\end{equation}
Next, the time derivative of the energy~\eqref{energy_in_terms_x} becomes:
\begin{widetext}
\begin{eqnarray}\label{time_derivative_energy}
    \frac{dE}{dt}&=&-\mu c^2 x \dot{x}\bigg[ 1-\frac{1}{2}\left( 3+\frac{1}{3}\nu \right)x^2-\frac{14}{9}\frac{\Lambda G^2 m^2}{c^4 x^6}+\frac{\Lambda G^2 m^2}{18 c^4 x^4}(103+95\nu) \nonumber \\
    && - \frac{33}{2}\frac{\Lambda G^2 m^2}{c^4 x^4}(1-3\nu)\mathrm{sin}^{2}(2\phi_{0\mathrm{PN}}) +O(x^4, \Lambda c^{-4}x^{-2}, \Lambda^2) \bigg] \nonumber\\
    && - \frac{11}{2}\mu\Lambda Gmcx(1-3\nu)\mathrm{sin}(4\phi_{0\mathrm{PN}}) +O(\Lambda x^2, \Lambda^2) \,,
\end{eqnarray}
\end{widetext}
%
%
where the PN parameter $x$ was used to express the frequency as $\dot{\phi}=\omega=\frac{c^3 x^3}{Gm}$. We equate the formulas~\eqref{power_in_term_x} and \eqref{time_derivative_energy} leading to the following expression to solve for the unknown PN parameter $x$, that is:
\begin{widetext}
\begin{eqnarray}
&& \int \mu c^2 x\bigg[ 1-\frac{1}{2}\left( 3+\frac{1}{3}\nu
 \right)x^2-\frac{14}{9}\frac{\Lambda G^2 m^2}{c^4 x^6}+\frac{\Lambda G^2 m^2}{18c^4 x^4}\left( 103+95\nu \right)-\frac{33\Lambda G^2 m^2}{2c^2 x^4}(1-3\nu)\mathrm{sin}^2 (2\phi_{0\mathrm{PN}})\bigg] \times \nonumber \\
 &&\bigg \lbrace \frac{32}{5}\frac{c^5}{G}\nu^2 x^{10}\bigg[ 1-\left( \frac{1247}{336}+\frac{35}{12}\nu \right)x^2-\frac{\Lambda G^2 m^2}{c^4 x^6}-\frac{\Lambda G^2 m^2}{432c^4 x^4}\left( 97-2692\nu \right)\bigg]\nonumber \\
 &+&\frac{11\mu \Lambda Gmcx}{2}(1-3\nu)\mathrm{sin}(4\phi_{0\mathrm{PN}})\bigg \rbrace^{-1}dx =-(t_c-t) \,.
\end{eqnarray}
\end{widetext}
%
%
Expanding in Taylor series we arrive at the following expression:
\begin{eqnarray}\label{Theta_eq}
    \Theta(t)&=& \frac{1}{256}x^{-8}\bigg[ 1+\frac{256}{192}\left( \frac{743}{336}+\frac{11}{4}\nu \right)x^2 \nonumber\\
    && \hspace{-1cm} - \frac{20}{63}\frac{\Lambda G^2 m^2}{c^4}x^{-6}-\frac{1}{54}\frac{\Lambda G^2 m^2}{c^4}(573+2444\nu)x^{-4} \nonumber \\
    && \hspace{-1cm} + \frac{132 \Lambda G^2 m^2}{c^4}(1-3\nu)x^8\int \frac{1}{x^{13}}\mathrm{sin}(2\phi_{0\mathrm{PN}})dx \nonumber\\
    && \hspace{-1cm}+O(x^4, \Lambda c^{-4}x^{-2}, \Lambda^2)
 \bigg] \,,
\end{eqnarray}
where $\Theta(t):= \frac{c^3\nu}{5Gm}(t_c-t)$, and $t_c$ is the time of coalescence. The inversion of the later equation reads:
\begin{eqnarray}\label{x_in_terms_theta}
    x&=& \frac{1}{2}\Theta^{-1/8}+\left( \frac{743}{16128}+\frac{11}{192} \nu\right)\Theta^{-3/8} \nonumber\\
    &-&\frac{80}{63}\frac{\Lambda G^2 m^2}{c^4} \Theta^{5/8}-\frac{1}{54}\frac{\Lambda G^2 m^2}{c^4}(573+2444\nu)\Theta^{3/8}\nonumber \\
    &+&\frac{33}{1024}\frac{\Lambda G^2 m^2}{c^4}(1-3\nu)\Theta^{-9/8}I(\Theta) \,,
\end{eqnarray}
with $I(\Theta):= \int \frac{1}{x^{13}}\mathrm{sin}(2\phi_{0\mathrm{PN}})dx$ (see Appendix~\ref{integral_I_Theta} for the explicit calculation). Note that the precise solution is a complex function; however, given or next numerical examples, only the real part is considered since its imaginary upshot is very small compared to its real counterpart. 
\subsection{Computation of the phase of oscillation of the GW of a two body compact system}
\begin{figure}
\centering
\includegraphics[width=0.45\textwidth]{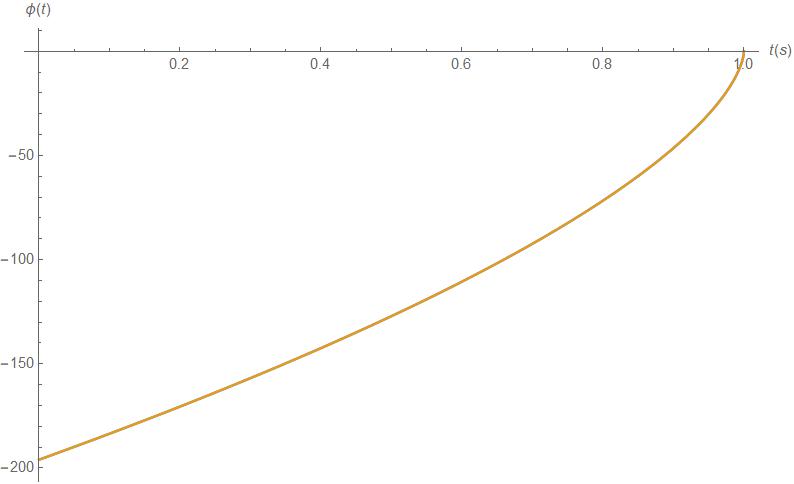}
\caption{Plot of the GW Newtonian phase $\phi_{0\mathrm{PN}}(t)$~\eqref{Newtonian_phase_time} for a binary compact system of identical masses with $\mathrm{m}=10^{31}\mathrm{kg}$, $\phi_0=0$, $\Lambda=10^{-52}\mathrm{m}^{-2}$, and $t_c=1s$. The blue line includes $\Lambda$, while the orange one does not ($\Lambda=0$). Note that both lines are superimposed on each other; thus, the effect of $\Lambda$ is negligible.}
\label{fig:4}
\end{figure}

First, to compute the phase of oscillation $\phi$ we know that
\begin{equation}
\frac{d\phi}{dt} = \frac{d\phi}{d\Theta}\frac{d\Theta}{dt} = -\frac{c^3 \nu}{5Gm}\frac{d\phi}{d\Theta} \,,  
\end{equation}
%
%
therefore we have
\begin{eqnarray}\label{phi_in_terms_x}
    \frac{d\phi}{d\Theta}&=& -\frac{5Gm}{c^3 \nu}\frac{d\phi}{dt} = -\frac{5Gm}{c^3\nu}\omega \nonumber \\
    &=& -\frac{5Gm}{c^3 \nu}\frac{c^3 x^3}{Gm} = -\frac{5}{\nu}x^3 \,,
\end{eqnarray}
%
%
where we have used the relation $\omega=\frac{c^3 x^3}{Gm}$. Moreover, from \eqref{x_in_terms_theta} we obtain:
\begin{eqnarray}
    x^3&=& \frac{1}{8}\Theta^{-3/8}+\left( \frac{743}{21504}+\frac{11}{256}\nu \right)\Theta^{-5/8} \nonumber\\
    &-&\frac{20}{21}\frac{\Lambda G^2 m^2}{c^4}\Theta^{3/8}-\frac{1}{72}\frac{\Lambda G^2 m^2}{c^4}(572+2444\nu)\Theta^{1/8} \nonumber \\
    &+&\frac{99}{4096}(1-3\nu)\frac{\Lambda G^2 m^2}{c^4}\Theta^{-11/8}I(\Theta) \,.
\end{eqnarray}
Finally, we integrate equation~\eqref{phi_in_terms_x} resulting in the following expression:
\begin{eqnarray}\label{phase_GW}
\hspace{-1cm} \phi(t)&=&\phi_0-\frac{1}{\nu}\bigg[ \Theta^{5/8}+\left( \frac{3715}{8064}+\frac{55}{96}\nu \right)\Theta^{3/8} \nonumber\\
    &-&\frac{800}{231}\frac{\Lambda G^2 m^2}{c^4}\Theta^{11/8} \nonumber\\
    &-&\frac{5}{81}\frac{\Lambda G^2 m^2}{c^4}(572+2444\nu)\Theta^{9/8} \nonumber \\
    &+&\frac{495}{4096}(1-3\nu)\frac{\Lambda G^2 m^2}{c^4}\int \Theta^{-11/8}I(\Theta)d\Theta \bigg] \,,
\end{eqnarray}
where $\phi_{0}$ is the value of the phase at the instant of coalescence.
First, note that in the limit $\Lambda \rightarrow 0$ equation~\eqref{phase_GW} matches to the known phase of the waveforms propagation of a GW~\cite{Luc_Blanchet, PhysRevD.51.5360}. Second, we present the figure~\ref{fig:4}, which is the graphic representation of the Newtonian phase $\phi_{0\mathrm{PN}}(t)$:
\begin{equation}\label{Newtonian_phase_time}
    \phi_{0\mathrm{PN}}=-\frac{5}{\nu}\bigg[ \frac{1}{5}\Theta^{5/8}-\frac{160}{231}\frac{\Lambda G^2 m^2}{c^4}\Theta^{11/8} \bigg],
\end{equation}
where we consider the binary compact system with both identical masses, such $\mathrm{m}=10^{31}\mathrm{kg}$; and $\phi_0=0$, $\Lambda=10^{-52}\mathrm{m}^{-2}$~\cite{Planck:2018nkj}, and $t_c=1s$. The orange line includes $\Lambda$, while the blue one does not ($\Lambda=0$). Note that both lines are superimposed on each other; thus, the effect of $\Lambda$ is negligible. Lastly, we observe that at $1$PN order, that is equation~\eqref{phase_GW}, the behavior of $\phi(t)$ due to $\Lambda$ is not modified whatsoever.  

%

%

\section{Gravitational waveforms in the time domain}\label{GW-hplus_hcross}
\begin{figure}
 \centering
 \includegraphics[width=0.5\textwidth]{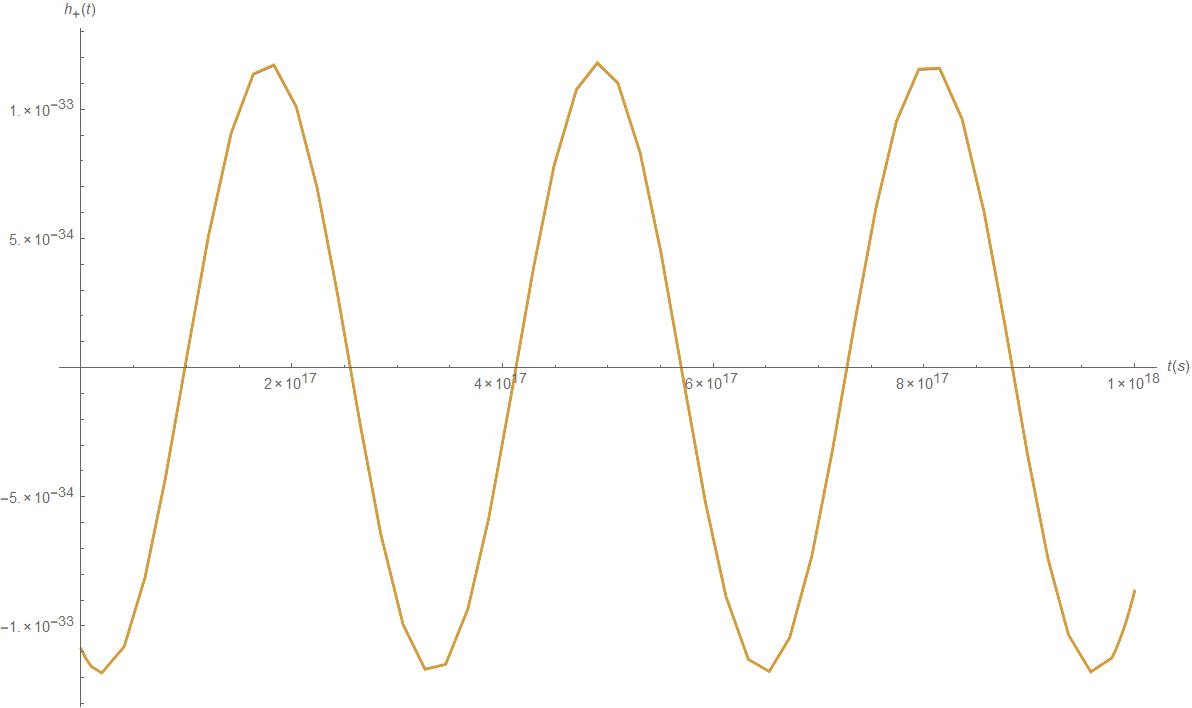}
 \includegraphics[width=0.5\textwidth]{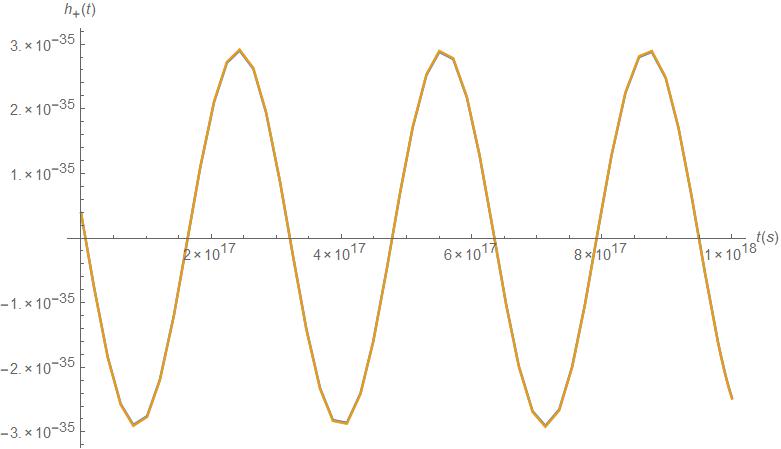}
 \caption{Plots of the gravitational waveform $h_{+}$ (top figure) and $h_{\times}$ (bottom figure) for a binary compact system of identical masses at $1$PN order with parameter values $m=10^{31}$kg, $R=200\times 10^{22}$m, $\omega=10^{-17}\mathrm{s}^{-1}$, $\Lambda=10^{-52}\mathrm{m}^{-2}$ and the inclination angle $\iota=\pi/2$ (top figure), $\iota=0$ (bottom figure). The blue line includes $\Lambda$, while the orange one does not ($\Lambda=0$). Note that two lines, in both polarizations, are superimposed on each other; thus, the effect of $\Lambda$ is negligible.}
 \label{fig:1}
\end{figure}

\begin{figure}
    \centering
    \includegraphics[width=0.5\textwidth]{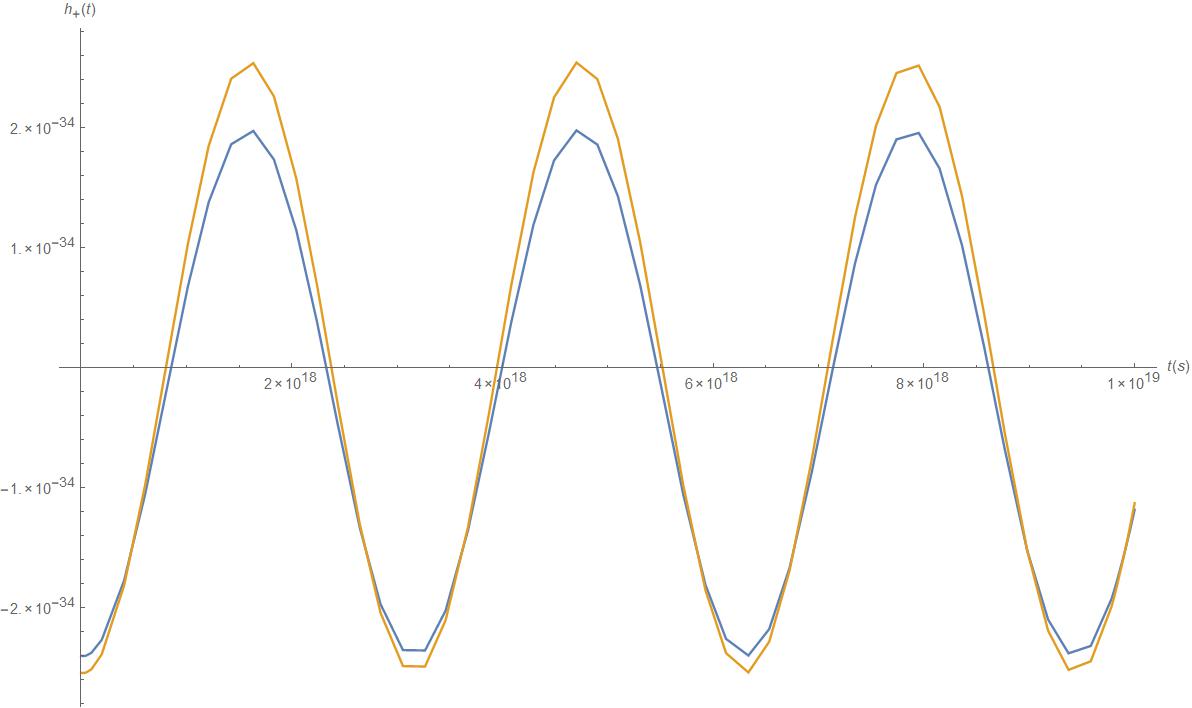}
    \includegraphics[width=0.5\textwidth]{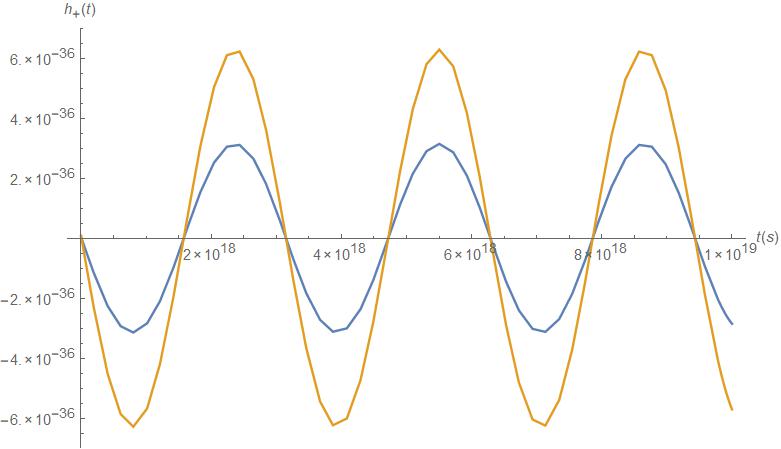}
    \caption{Plots of the gravitational waveforms $h_{+}$ (top figure) and $h_{\times}$ (bottom figure) for a binary compact system of identical masses at $1$PN order. The parameters are given by $m=10^{31}$Kg, $R=200\times 10^{22}$m, $\omega=10^{-18}\mathrm{s}^{-1}$ and the inclination angle $\iota=\pi/2$ (top figure), $\iota=0$ (bottom figure). The blue line includes $\Lambda$, while the orange one does not ($\Lambda=0$). Note that with this particular frequency, the effect of $\Lambda$ starts to be observable.} 
    \label{fig:2}
\end{figure}

\begin{figure}
    \centering
    \includegraphics[width=0.5\textwidth]{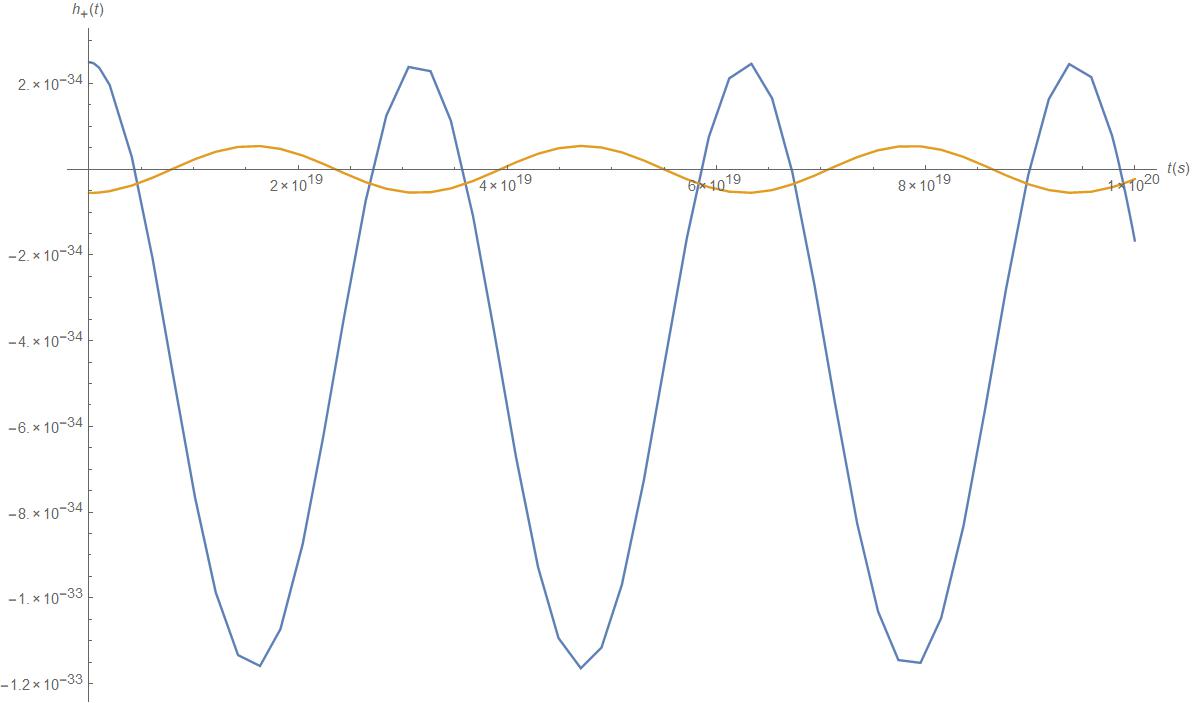}
    \includegraphics[width=0.5\textwidth]{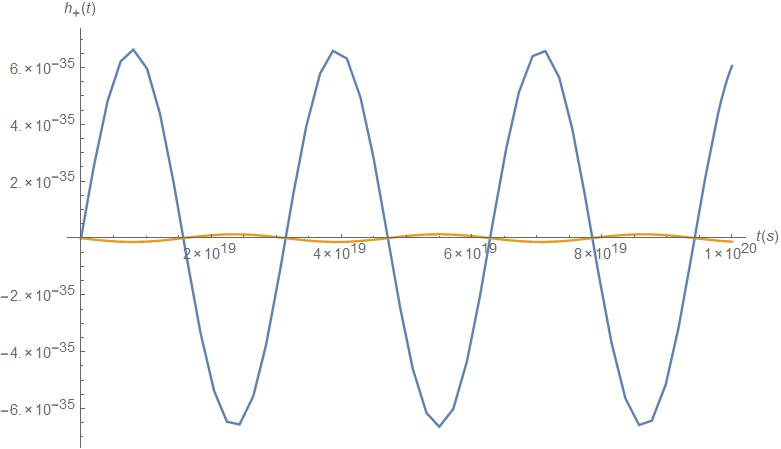}
    \caption{Plots of the waveforms $h_{+}$ (top figure) and $h_{\times}$ (bottom figure) for a binary compact system of identical masses at $1$PN order. The parameter values are given by $m=10^{31}$kg, $R=200\times 10^{22}$m, $\omega=10^{-19}\mathrm{s}^{-1}$ with inclination angle $\iota=\pi/2$ (top figure), $\iota=0$ (bottom figure). The blue line includes $\Lambda$, while the orange one does not ($\Lambda=0$). Note that with this particular frequency, the effect of $\Lambda$ becomes very evident.}
    \label{fig:3}
\end{figure}

To compute the time domain we must introduce the orthonormal triad $\hat{N},\hat{p}$, and $\hat{q}$; with $\hat{N}$ as the unit vector, which is a radial vector pointing from the source to the observer, $\hat{p}$ lies on the intersection of the orbital plane with the plane of the sky and $\hat{q}=\hat{N}\times \vec{p}$. We also have to consider the parameters $\iota$ and $\phi$ that are the inclination angle relative to $\hat{N}$ and the orbital phase of the motion of the body $1$ measured in positive sense from the line of nodes; and $\hat{n}$ is the unitary vector of $\Vec{r}$. Thus, we have that
\begin{eqnarray}
   \label{p_orbital} \hat{p}&=& (1,0,0) \,, \\
   \label{q_orbital} \hat{q}&=& (0,\mathrm{cos}\iota, -\mathrm{sin}\iota) \,, \\
   \label{N_orbital} \hat{n}&=& \hat{p}\mathrm{cos}\phi+(\hat{q}\mathrm{cos}\iota+\hat{N}\mathrm{sin}\iota)\mathrm{cos}\phi, \\
    \label{lambda_orbital}\hat{\lambda}&=& -\hat{p}\mathrm{sin}\phi+(\hat{q}\mathrm{cos}\iota+\hat{n}\mathrm{sin}\iota)\mathrm{cos}\phi\,,  
\end{eqnarray}
where $\vec{v}=rw\hat{\lambda}$ for circular orbits. The gravitational waveforms in the time domain are linear combinations of the polarizations waveforms $h_{+}(t)$ and $h_{\times}(t)$ defined by the projections:
\begin{eqnarray}
   \label{h+} h_{+}&=&\frac{1}{2}(\hat{p}_i \hat{p}_j-\hat{q}_i\hat{q}_j)h^{ij} \,, \\
    \label{hx} h_{\times}&=& \frac{1}{2}(\hat{p}_i\hat{q}_j+\hat{q}_i\hat{p}_j)h^{ij} \,.
\end{eqnarray}
We have already computed the waveforms extracted after applying their projections (see~\eqref{wave_form_complete}), and recall that we have taken the particular case for a circular motion $\dot{r}=0$, therefore both polarizations become: 
%
\begin{eqnarray}
     \label{h+waveform} h_{+}&=& \frac{2G\mu}{c^2 R}\bigg( \frac{Gm\omega}{c^3} \bigg)^{2/3}\bigg\lbrace H_{+}^0+xH_{+}^{1/2}+x^2 H_{+}^1 \nonumber\\
    && + O(x^3, \Lambda c^{-1},\Lambda^2)\bigg\rbrace \,, \\
    \label{h_x_waveform} h_{\times}&=& \frac{2G\mu}{c^2 R}\bigg( \frac{Gm\omega}{c^3} \bigg)^{2/3}\bigg\lbrace H_{\times}^0+xH_{\times}^{1/2}+x^2 H_{\times}^{1} \nonumber\\
    && + O(x^3, \Lambda c^{-1}, \Lambda^2) \bigg\rbrace \,,
\end{eqnarray}
with
\begin{widetext}
\begin{eqnarray}
\label{H_+0} H_{+}^0&=& -(1+\mathrm{cos}^2\iota)\mathrm{cos}2\phi+\frac{\Lambda c^2}{\omega^2}\left(  -\frac{1}{12}\mathrm{sin}^2 \iota+\frac{5}{36}(1+\mathrm{cos}^2\iota)\mathrm{cos}2\phi \right) \,, \\
\label{H_+12} H_{+}^{1/2}&=& -\frac{\Delta m}{m}\frac{1}{8}\mathrm{sin}\iota  \bigg[ (5+\mathrm{cos}^2 \iota)\mathrm{cos}\phi -9(1+\mathrm{cos}^2\iota)\mathrm{cos}3\phi \bigg]\bigg( 1-\frac{\Lambda c^2}{3\omega^2} \bigg) \,, \\
\label{H_+1} H_{+}^1&=& \frac{1}{6}\lbrace [19+9\mathrm{cos}^2\iota-2\mathrm{cos}^4\iota]-\nu[19-11\mathrm{cos}^2\iota-6\mathrm{cos}^4\iota] \rbrace \mathrm{cos}2\phi  \nonumber\\
        && - \frac{4}{3}\mathrm{sin}^2\iota(1+\mathrm{cos}^2\iota)(1-3\nu)\mathrm{cos}4\phi  \nonumber\\
        && + \frac{\Lambda c^2}{\omega^2} \left\lbrace \frac{13}{24}-\frac{9}{16}\mathrm{cos}^2\iota+\frac{1}{48}\mathrm{cos}^4\iota+\frac{275}{72}\nu \mathrm{sin}^2\iota+\mathrm{cos}2\phi \bigg[ -\frac{371}{432}-\frac{35}{144}\mathrm{cos}^2\iota -\frac{35}{108}\mathrm{cos}^4\iota \right. \nonumber\\
        && + \nu\bigg( \frac{331}{144}+\frac{65}{144}\mathrm{cos}^2\iota+\frac{35}{36}\mathrm{cos}^4\iota \bigg) \bigg] \nonumber\\
        && + \left. \mathrm{cos}4\phi\bigg[ \frac{5}{18}+\frac{11}{54}\mathrm{cos}^2\iota-\frac{13}{27}\mathrm{cos}^4\iota+\nu \bigg( -\frac{5}{6}-\frac{69}{72}\mathrm{cos}^2\iota+\frac{13}{9}\mathrm{cos}^4\iota \bigg) \bigg] \right\rbrace \,, \\
\label{H_x0} H_{\times}^0&=& -2\mathrm{cos}\iota \mathrm{sin}2\phi +\frac{\Lambda c^2}{9\omega^2}\mathrm{cos}\iota \mathrm{sin}2\phi \,, \\
\label{H_x12}  H_{\times}^{1/2}&=& -\frac{\Delta m}{m}\frac{3}{8}\mathrm{sin}2\iota  \bigg[ \bigg( 1+\frac{2}{9}\frac{\Lambda c^2}{\omega^2} \bigg)\mathrm{sin}\phi -\bigg( 3-\frac{20}{9}\frac{\Lambda c^2}{\omega^2}
 \bigg)\mathrm{sin}3\phi \bigg] \,, \\
\label{H_x1}  H_{\times}^{1}&=&\mathrm{cos}\iota \bigg[ \left \lbrace  \bigg( \frac{17}{3}-\frac{4}{3}\mathrm{cos}^2\iota \bigg)+\nu \bigg( -\frac{13}{3}+4\mathrm{cos}^2\iota \bigg)  \right \rbrace \mathrm{sin}2\phi -\frac{8}{3}(1-3\nu)\mathrm{sin}^2\iota \mathrm{sin}4\phi \nonumber\\
 && + \frac{\Lambda c^2}{\omega^2} \left \lbrace  \bigg( -\frac{92}{27}+\frac{1}{3}\mathrm{cos}^2\iota \bigg)+\nu \bigg( \frac{79}{18}-\frac{13}{6}\mathrm{cos}^2\iota
 \bigg)  \right \rbrace \mathrm{sin}2\phi +\frac{\Lambda c^2}{\omega^2}   \bigg( \frac{359}{216}-\frac{359}{72}\nu \bigg)\mathrm{sin}^2\iota   \mathrm{sin}4\phi \bigg] \,.
\end{eqnarray}
\end{widetext}
The following identities, which come from the combinations of the definitions \eqref{h+} and \eqref{hx} with \eqref{p_orbital}-\eqref{lambda_orbital}, were utilized to compute of the above polarizations $h_{+}\,, h_{\times}$: 
 \\
 \\
 \\
\begin{eqnarray}
  (\hat{n}^i\hat{n}^j)_{+}&=& \frac{1}{4}\mathrm{sin}^2 \iota +\frac{1}{4}[1+\mathrm{cos}^2\iota]\mathrm{cos}2\phi \,, \\
    (\hat{\lambda}^i \hat{\lambda}^j)_{\times}&=& -\frac{1}{2}\mathrm{cos}\iota \mathrm{sin}2\phi \,, \\
    (\hat{n}^{(i}\hat{\lambda}^{j)})_{+}&=&-\frac{1}{4}[1+\mathrm{cos}^2\iota]\mathrm{sin}2\phi \,, \\
    (\hat{n}^i \hat{n}^j)_{\times}&=& \frac{1}{2}\mathrm{cos}\iota \mathrm{sin}2\phi \,, \\
    (\hat{\lambda}^i \hat{\lambda}^j)_{+}&=& \frac{1}{4}\mathrm{sin}^2\iota-\frac{1}{4}[1+\mathrm{cos}^2\iota]\mathrm{cos}2\phi \,, \\
    (\hat{n}^{(i}\hat{\lambda}^{j)})_{\times}&=& \frac{1}{2}\mathrm{cos}\iota \mathrm{cos}2\phi \,, \\
    \left( (r_i)^2 r^{(i}r^{j)} \right)_{\times}&=& \frac{1}{4}r^4 \mathrm{cos}\iota \mathrm{sin}2\phi \,, \\
    \left( (r_i)^2 r^{(i}r^{j)} \right)_{+}&=& \frac{1}{16}r^4 [ 3+5\mathrm{cos}2\phi \nonumber\\
    &-&3\mathrm{cos}^2\iota (1-\mathrm{cos}2\phi)] \,, \\
 \left[ (v_i)^2r^{(i}r^{j)} \right]_{\times}&=& \left[ r_i v_i v^{(i}r^{j)} \right]_{\times} \nonumber\\
 &=& \frac{1}{4}r^4 \omega^2 \mathrm{cos}\iota \mathrm{sin}2\phi \,, \\
 \left[ (v_i)^2r^{(i}r^{j)} \right]_{+}&=& \left[ r_i v_i v^{(i}r^{j)} \right]_{+} \nonumber\\
 &=& \frac{1}{16}r^4 \omega^2 \mathrm{sin}^2 \iota (1-\mathrm{cos}4\phi) \,, \\
    \hat{N}\cdot \hat{n}&=& \mathrm{sin}\iota \mathrm{sin}\phi \,, \\
    \hat{N}\cdot \hat{\lambda}&=& \mathrm{sin}\iota \mathrm{cos}\phi \,,
\end{eqnarray}
here all repeated indices do not indicate sum. 
From equations \eqref{H_+0}-\eqref{H_x1} one can observe that all $H_{i}$'s present nearly the same structure, that is, they have a constant term multiplied by a trigonometry function, which contains the phase $\phi$; except all $H_{+}$'s with $\Lambda$, they also present a shift constant. As a consequence of this fact, we can say that with $\Lambda$, at constant frequencies $\phi=\omega(t-R/c)$, 
the amplitude of the waveforms change in magnitude and their roots (points where the function vanishes) are modified only in the $h_{+}$ polarization (see figure~\ref{fig:2} and figure~\ref{fig:3}). We close this section with four remarks:


i) The first case presented in figure~\ref{fig:1} with $\omega = 10^{-17}\mathrm{s}^{-1}$ one can see that the two lines (blue with $\Lambda\simeq 10^{-52}\mathrm{m}^{-2}$~\cite{Planck:2018nkj}; orange $\Lambda=0$) in both polarizations $h_{+}$ and $h_{\times}$ are superimposed on each other; thus, the effect of $\Lambda$ is negligible. 


ii) In the second case with $\omega = 10^{-18}\mathrm{s}^{-1}$ (see figure~\ref{fig:2}) we can notice that $\Lambda$ begins to have importance. We observe that $\Lambda$ modifies the amplitudes of $h_{+}$ and $h_{\times}$, it reduces their sizes compare to the case with null $\Lambda$. This is a direct consequence that the cosmological constant ``stretches" the spacetime making that the objects within it move away from each other.


iii) For the particular frequencies $\omega_{0}=c\sqrt{5\Lambda}/6$ and $\omega_0=c\sqrt{2\Lambda}/6$, the amplitudes of $h_{+}$ and $h_{\times}$ are canceled respectively, at $0$PN order. Thus, if the system oscillates at one of this particular frequencies, $\Lambda$ would annihilate such amplitudes at Newtonian order. Nevertheless, we observe that the spacetime is still altered by $\Lambda$ in an oscillatory way since the correction of the waveforms at $0.5$PN order or higher, in fact, prevail (see \eqref{h+waveform} and \eqref{h_x_waveform}).


iv) Then, for $\omega < \omega_0$, as shown in figure~\ref{fig:3} the effect of $\Lambda$ becomes very evident. The ripples of the spacetime  are now ``stretched" by the cosmological constant. In fact, one can drop all the expressions without $\Lambda$ from \eqref{h+waveform} and \eqref{h_x_waveform}, and we will get nearly the same output; therefore, the waveforms of the GW depend mostly on those terms with $\Lambda\neq 0$. Furthermore, observe that in the case of the plot of $h_{+}$ the crest and trough are displaced downwards as a consequence of the shift constants; such as the coefficient $-c^{2}\mathrm{sin}^2\iota/(12\omega^{2})$ that multiplies $\Lambda$ in \eqref{H_+0}. 


On the other hand, there is an exception among all aforementioned examples, that is the case iii). There is no difference between the plots of $0$PN and $1$PN orders due to the very small value of the frequency $\omega$. The corresponding terms at $1$PN are practically negligible in comparison to the $0$PN ones. Nonetheless, for the case of higher frequencies; i.e., $\omega\geq 10^{18} \mathrm{s}^{-1}$ there might be a difference between results at $0$PN and $1$PN but the presence of $\Lambda$ is negligible for numerical purposes. Hence we can confirm that at higher orders of the Post-Newtonian method, the presence of $\Lambda$ will not affect the polarizations $h_{+}$ and $h_{\times}$.

%
\section{Concluding remarks}\label{Sec_conclusions}
%
In this paper we have studied from scratch the propagation of GWs including the cosmological constant $\Lambda$ in a binary compact system. Using the direct integration of the relaxed EFE at 1PN, and taking into account that the terms $O(\Lambda h)$ were dropped given that, from the beginning, we assume that $\Lambda \simeq10^{-52}\mathrm{m}^{-2}$~\cite{Planck:2018nkj} is very small and positive~\cite{PhysRevD.81.084002}. We also compute the wave forms~\eqref{wave_form_complete}, where the equations of motion at 1PN~\eqref{EOM} were derived from the Lagrangian taken at the center of mass frame, expressed in~\eqref{two_body_lagrangian_CM}. Furthermore, observing the solutions for $h^{00}$ and $\overset{(2)}{g}_{00}$ given by \eqref{h00} and \eqref{g_00_solution}, correspondingly; we find that $\Lambda$ can be interpreted as a PN factor since globally we can factorize $1/c^2$ and this power of $c$ is the $1$PN approximation. 

Focusing on the particular case of a binary quasicircular motion, we derive the energy and the radiated power given by \eqref{energy_two_body} and \eqref{radiated_power_1}, respectively. Then, we substitute these results into the balance equation \eqref{balance_equation}, where the PN parameters $\gamma$ and $x$ were introduced, in order to obtain the phase \eqref{phase_GW} in the time domain at $1$PN order. We can notice that this expression depends explicitly of their own quasicircular phase orbit $\phi(t)$; nevertheless, we can use the Newtonian phase \eqref{Newtonian_phase} to compute the integral~\eqref{result_integral_I_Theta} (given in Appendix \ref{integral_I_Theta}). On the other hand, from figure~\ref{fig:4} we can observe that $\phi$ behaves the same with or without $\Lambda$; therefore, adding the cosmological constant does not affect the phase. However, the impact of $\Lambda$ starts becoming noticeable on the amplitudes of the polarizations $h_{+}$ and $h_{\times}$ (see Figures \ref{fig:2} and \ref{fig:3}) when taking a constant frequency $\omega<10^{-18}\rm s^{-1}$. Moreover, we find that given the particular frequencies $\omega_{0}=c\sqrt{5\Lambda}/6$ and $\omega_{0}=c\sqrt{2\Lambda}/6$ the amplitudes of $h_{+}$~\eqref{h+waveform} and $h_{\times}$~\eqref{h_x_waveform} vanish at Newtonian order; nonetheless, at higher orders the propagation of the GWs holds.

In the near future, we can extend our study now considering $O(\Lambda h)$ terms~\cite{PhysRevD.84.063523}, given that in the early stages of the universe (inflationary period~\cite{PhysRevD.23.347,LINDE1982389,PhysRevLett.48.1220}) the value of $\Lambda$ could have been much larger. Also, we may investigate heavier objects, such as a system of black holes at the center of two galaxies weighing billions of solar masses, since they emit GWs with lower frequencies. This, indeed, opens the possibility to explore detectable signals from the most recent NANOGrav survey~\cite{NANOGrav:2023gor}. Furthermore, complementing our work applied to the scalar-Gauss-Bonnet-gravity~\cite{Shiralilou_2022} could shed light to understand the behavior of $\Lambda$ together with the scalar field. On the other hand, we can explore the coordinate transformation from the Cartesian coordinates given in the spatial components of \eqref{matrix_form} which leads to the Shwarzschild-de-Sitter metric~\cite{ PhysRevD.81.084002} (see also~\cite{PhysRevD.84.063523,Laszlo}). We can also expand this PN approach from the very beginning in the Brans-Dicke theory~\cite{Brans-Dicke-Cosmological} (see also~\cite{Bernard_2022}). 

\acknowledgments
This work was supported by the CONAHCYT Network Project No. 376127 {\it Sombras, lentes y ondas gravitatorias generadas por objetos compactos astrofísicos}. R.H.J and R.E. are supported by CONAHCYT Estancias Posdoctorales por M\'{e}xico, Modalidad 1: Estancia Posdoctoral Acad\'{e}mica and by SNI-CONAHCYT. C.M. wants to thank SNI-CONAHCYT and PROSNI-UDG.

\appendix

\section{Metric at Newtonian order}\label{Metric_Newtonian_order}
%
This appendix is devoted to compute the components of the metric at Newtonian order. We follow the method developed in~\cite{10.1093/oso/9780198786399.001.0001}. To begin, we make the expansion of the metric in the PN approximation as follows:
\begin{eqnarray}
    g_{00}&=& -1+\overset{(2)}{g}_{00}+\cdots \\
    g_{0i}&=& \overset{(3)}{g}_{0i}+ \cdots \\
    g_{ij}&=& \delta_{ij}+\overset{(2)}{g}_{ij}+\cdots 
\end{eqnarray}
where the number over the objects means the power of the factor of the velocity ratio $v/c$. The temporal and spatial components of the Ricci tensor take the following form:
\begin{eqnarray}
    \overset{(2)}{R}_{00}&=&-\frac{1}{2}\nabla^2 \overset{(2)}{g}_{00}, \\
    \overset{(2)}{R}_{ij}&=& \frac{1}{2}\left[ \partial_i \left( \frac{1}{2}\partial_j \overset{(2)}{g}_{00}-\frac{1}{2}\partial_j\overset{(2)}{g}_{kk}+\partial_k \overset{(2)}{g}_{jk} \right) \right. \nonumber\\
    &+&\left. \partial_j \left( \frac{1}{2}\partial_i \overset{(2)}{g}_{00}-\frac{1}{2}\partial_i \overset{(2)}{g}_{kk}+\partial_k \overset{(2)}{g}_{ik} \right)-\nabla^2 \overset{(2)}{g}_{ij} \right] \nonumber \\
    &=& \frac{1}{2}\left[ \partial_i \Gamma_j +\partial_j \Gamma_i -\nabla^2 \overset{(2)}{g}_{ij} \right] \,,
\end{eqnarray}
where we define $\Gamma_i:=\frac{1}{2}\partial_i \overset{(2)}{g}_{00}-\frac{1}{2}\partial_i\overset{(2)}{g}_{kk}+\partial_k \overset{(2)}{g}_{ik}$. Since we are considering a system of $n$ compact bodies, the expression energy-momentum tensor that describes it is given by:
\begin{equation}\label{EM_tensor_two_bodies}
    T^{\mu\nu}= \frac{1}{\sqrt{-g}}\sum_a m_a \frac{d\tau_a}{dt}\frac{dx_a^{\mu}}{d\tau_a}\frac{dx_a^{\nu}}{d\tau_a}\delta^3 (\Vec{x}-\Vec{x}_a(t)) \,,
\end{equation}
with $\tau_a$ as the proper time of the particle $a$ and $t$ as the time coordinate. Furthermore, the EFE \eqref{EFE} can be rewritten as follows:
\begin{eqnarray}\label{EFE_2}
    R_{\mu\nu}&=&\frac{8\pi G}{c^4}(T_{\mu\nu}-\frac{1}{2}Tg_{\mu\nu})+\Lambda g_{\mu\nu} \nonumber\\
    &:=&S_{\mu\nu} \,,
\end{eqnarray}
with $S_{\mu\nu}$ playing the role as the source of the EFE. The sources components $S_{00}$ and $S_{ij}$ at lowest order correspondingly read:
\begin{eqnarray}
    S_{00}&=&\frac{4\pi G}{c^4}\overset{(0)}{T}_{00}-\Lambda, \\
    S_{ij}&=&\frac{4\pi G}{c^4}\delta_{ij}\overset{(0)}{T}_{00}+\Lambda \delta_{ij} \,,
\end{eqnarray}
where we use the result $\overset{(0)}{T}_{00}\simeq -\overset{(0)}{T}$. From the EFE we find the following set of equations:
\begin{eqnarray}
   \label{set_00} -\frac{1}{2}\nabla^2 \overset{(2)}{g}_{00}&=& \frac{4\pi G}{c^4}\overset{(0)}{T}_{00}-\Lambda, \\
    \label{set_ij}\frac{1}{2}\left[ \partial_i \Gamma_j +\partial_j \Gamma_i -\nabla^2 \overset{(2)}{g}_{ij} \right]&=&\frac{4\pi G}{c^4}\delta_{ij}\overset{(0)}{T}_{00}+\Lambda \delta_{ij} \,.
\end{eqnarray}
Recalling that $\overset{(0)}{T}_{00}=\sum_a m_a\delta^3 (\vec{x}-\vec{x}_a(t))$, the solution of \eqref{set_00} is showed as:
\begin{equation}\label{g_00_solution}
    \overset{(2)}{g}_{00}=\frac{2G}{c^2}\sum_a \frac{m_a}{r_a}+\frac{\Lambda}{3}|\vec{x}|^2 \,,
\end{equation}
with $\vec{x}$, $\vec{x}_a$ as the vector field point and the vector position of the particle $a$, respectively; and $r_a:=|\vec{x}-\vec{x}_a|$. On the other hand, to solve \eqref{set_ij}, it is convenient to choose the gauge $\Gamma_i=0$. Therefore, the solution is given by:
\begin{equation}\label{g_ij_solution}
    \overset{(2)}{g}_{ij}=\bigg[ \frac{2G}{c^2}\sum_a \frac{m_a}{r_a}-\frac{\Lambda}{2}\left( \frac{1}{3}|\Vec{x}|^2+x_i^2 \right) \bigg]\delta_{ij} \,.
\end{equation}
Notice that $\overset{(2)}{g}_{ij}$ is a diagonal matrix, but it is not proportional to the identity, and the solution \eqref{g_ij_solution} satisfies \eqref{set_ij} as long as the gauge\footnote{See~\cite{Weinberg:1972kfs} to notice that the gauge $\Gamma_i=0$ is equivalent to fix the spatial components of the De Donder gauge condition $\partial_\mu(\sqrt{-g}g^{\mu i})=0$, which is used in the linearized form of the wave form corresponding to $\partial_\mu h^{\mu i}=0$.} $\Gamma_i=0$ holds. Furthermore, observe that the solutions \eqref{g_00_solution} and \eqref{g_ij_solution} meet the gauge $\Gamma_i=0$ as well.
\section{Two body Lagrangian of a compact system}\label{two_lagrangian}
In this section, the interaction of a two body compact system at the first Post-Newtonian correction order with cosmological constant is considered. To begin with the analysis, we propose the following ansatz of the components of the metric as follows:
\begin{eqnarray}
    \label{g_00}g_{00}&=& -e^{2U}+O(c^{-4}, \Lambda c^{-2}) \,, \\
    \label{g_0i}g_{0i}&=& 4g_i+O(c^{-5}, \Lambda c^{-3}) \,, \\
    \label{g_ij}g_{ij}&=& \delta_{ij}e^{-[2U+\frac{\Lambda}{2}(|\vec{x}|^{2}+(x_i)^2)]}+O(c^{-4}, \Lambda c^{-2}) \,,
\end{eqnarray}
with no sum over the index $i$ in the last term of $g_{ij}$. Introducing the objects \eqref{g_00}, \eqref{g_0i}, and \eqref{g_ij} into the temporal and spatial-temporal components of the Ricci tensor we find:
\begin{eqnarray}
    \label{R_00}R_{00}&=&\nabla^2 U+\frac{1}{c}\partial_t \left( \frac{1}{c}3\partial_t U +4 \partial_i g_i+\frac{2}{c}\Lambda \Vec{x}\cdot \dot{\Vec{x}} \right) \,, \\
    \label{R_0i}R_{0i}&=& -2\nabla^2 g_{i} +2\partial_i \left[ \frac{1}{c}\dot{U}+\partial_j g_j +\frac{3}{4c}\Lambda (\Vec{x}\cdot \dot{\Vec{x}})\right. \nonumber \\
    &-&\left. \frac{\Lambda}{4c}(x_i \dot{x}_i) \right] \,,
\end{eqnarray}
again with no sum over the index $i$ in the last term of $R_{0i}$. Recalling that the source of the EFE can be defined from~\eqref{EFE_2}, we find that at Newtonian order; namely when the factor $v/c\rightarrow 0$,
\begin{eqnarray}
   \label{S_00} S_{00}&=& \frac{4\pi G}{c^4} \left( T_{00}+T_{ii} \right)-\Lambda \,, \\
   \label{S_0i} S_{0i}&=& \frac{8\pi G}{c^4}T_{0i} \,.
\end{eqnarray}
Consequently, plugging back the components of the Ricci tensor
\eqref{R_00} and \eqref{R_0i}, and the components of the source \eqref{S_00} and \eqref{S_0i} into the EFE \eqref{EFE_2} yields:
\begin{widetext}
\begin{eqnarray}
    \label{U_eq}\nabla^2 U +\frac{1}{c}\partial_t \left( \frac{1}{c}3\partial_t U+4\partial_i g_i +\frac{2}{c}\Lambda \Vec{x}\cdot \dot{\Vec{x}} \right)&=&\frac{4\pi G}{c^4}\left( T_{00}+T_{ii} \right)+g_{00}\Lambda \,, \\
    \label{g_i_eq}-2\nabla^2 g_i +2\partial_i \left[ \frac{1}{c}\dot{U}+\partial_j g_j+\frac{3}{4c}\Lambda (\Vec{x}\cdot \dot{\Vec{x}})-\frac{\Lambda}{4c}(x_i \dot{x}_i) \right]&=& \frac{8\pi G}{c^4}T_{0i} \,,
\end{eqnarray}
\end{widetext}
with no sum in the index $i$ in the last term of the second equation. Using the energy-momentum tensor given in \eqref{EM_tensor_two_bodies}, the components of the matter sources showed in \eqref{U_eq} and \eqref{g_i_eq} become:
\begin{eqnarray}
   \label{T00} T^{00}&=& c^2 \sum_a m_a \bigg[ 1+U+\frac{v^2}{2c^2}\left( 1+\Lambda |\Vec{x}|^2 \right) \nonumber\\
   &+&  \Lambda |\Vec{x}|^2 \bigg]\delta^3 (\Vec{x}-\Vec{x}_a(t)) \,, \\
    T^{0i}&=& -c\sum_a m_a v_{ai}\delta^3(\vec{x}-\vec{x}_a(t)), \\
   \label{Tii} T^{ii}&=&\sum_a m_a v_a^2 \left[ 1+\Lambda |\Vec{x}|^2 \right] \delta^3 (\Vec{x}-\Vec{x}_a(t)) \,.
\end{eqnarray}
On the other hand, to solve \eqref{U_eq}, it is convenient to use the `Coulomb-like' gauge~\cite{10.1093/oso/9780198786399.001.0001} considering the presence of the cosmological constant as follows:
\begin{equation}\label{Coulomb_gauge}
    \frac{1}{c}3 \dot{U}+4 \partial_i g_i+\frac{2}{c}\Lambda \Vec{x}\cdot \dot{\Vec{x}}=0 \,.
\end{equation}
Therefore, regarding this particular gauge, \eqref{U_eq} takes the following form:
\begin{equation}\label{nabla_U}
    \nabla^2 U=\frac{4\pi G}{c^4} (T_{00}+T_{ii})+\Lambda g_{00} \,.
\end{equation}
Next, we define the object $\xi_i$ that satisfies the subsequent relation:
\begin{equation}\label{nabla_xi}
    \nabla^2 \xi_i=\nabla^2 g_i-\partial_i \left[ \frac{1}{c}\dot{U}+\partial_j g_j+\frac{3}{4c}\Lambda (\Vec{x}\cdot \dot{\Vec{x}})-\frac{\Lambda}{4c}(x_i \dot{x}_i)   \right] \,,
\end{equation}
with no sum in the $i$ index. And we define:
\begin{equation}\label{g_i_relation}
    g_{i}:= \xi_i +\frac{1}{4c}\partial_i \dot{\chi}+\frac{\Lambda}{18c} \partial_i\left( x^3 \dot{x}+y^3 \dot{y}+z^3 \Dot{z} \right) \,.
\end{equation}
Remarkably the addition of the last term of \eqref{g_i_relation}
ensures that the `Coulomb-like' gauge \eqref{Coulomb_gauge} holds. Moreover, the third term with $\Lambda$ has no rotational symmetry; hence, the `Coulomb-like' gauge breaks the rotational symmetry of the component of $g_{0i}$ at order $O(c^{-3})$.
%
%
We then apply the Laplacian operator to both sides of previous relation, having:
\begin{eqnarray}
    -\frac{1}{4c}\partial_i \nabla^2 \dot{\chi}&=& \nabla^2 \xi_i-\nabla^2 g_i +\frac{\Lambda}{3c}\partial_i(\vec{x}\cdot \Dot{\Vec{x}}) \nonumber \\
    &=& -\partial_i \left[ \frac{1}{c}\dot{U}+\partial_j g_j \right] \nonumber \\
    &&-\partial_i \left[ \frac{3}{4c}\Lambda (\Vec{x}\cdot \Vec{x})-\frac{1}{4c}\Lambda (x_i \dot{x}_i) \right] \nonumber \\
    &&+\frac{\Lambda}{3c}\partial_i (\vec{x}\cdot \Dot{\vec{x}}) \,,
\end{eqnarray}
with no sum in the index $i$; and where we have used~\eqref{nabla_xi}. Integrating the above relation we obtain:
\begin{eqnarray}
    -\frac{1}{4c}\nabla^2 \dot{\chi}&=&-\bigg[ \frac{1}{c}\dot{U}+\partial_j g_j+\frac{3\Lambda}{4c}(\vec{x}\cdot \dot{\vec{x}}) \nonumber\\
    &&-\frac{\Lambda}{4c}(x_i\dot{x}_i)-\frac{\Lambda}{3c}(\vec{x}\cdot \Dot{\Vec{x}}) \bigg] \,,
\end{eqnarray}
with no sum over the index $i$. Note that the last expression represents 3 equations given that $i=1,2,3$; and this fact is a direct consequence that the rotational symmetry is broken due to the presence of $\Lambda$. However, if we add them all together we obtain:
\begin{equation}\label{div_chi_dot}
    -\frac{1}{4c}\nabla^2 \dot{\chi}=-\left[ \frac{1}{c}\dot{U}+\partial_j g_j \right]-\frac{1}{3c}\Lambda (\Vec{x}\cdot \dot{\Vec{x}}) \,.
\end{equation}
Then, we apply the `Coulomb-like' gauge \eqref{Coulomb_gauge}, having:
\begin{equation}
    \frac{1}{4c}\nabla^2 \dot{\chi}= \partial_t \left[ \frac{1}{4c}U-\frac{\Lambda}{12c}|\Vec{x}|^2 \right] \,.
\end{equation}
Next, we integrate with respect to time both sides:
\begin{equation}\label{nabla_chi}
    \nabla^2 \chi = U-\frac{\Lambda}{3}|\Vec{x}|^2 \,.
\end{equation}
On the other hand, replacing \eqref{nabla_xi} in \eqref{g_i_eq} yields:
\begin{equation}\label{nabla_xi_i}
    \nabla^2 \xi_i= -\frac{4\pi G}{c^4}T_{0i} \,.
\end{equation}
Now, to solve \eqref{U_eq}, we begin solving it at lowest order. So, that relation becomes:
\begin{eqnarray}
    \nabla^2 U &=& \frac{4\pi G}{c^4}(T_{00}-T_{ii})-\Lambda \nonumber \\
    &\simeq& \frac{4\pi G}{c^2}\sum_a m_a \delta^3(\Vec{x}-\Vec{x}_a(t)) \nonumber \\
    &=& \nabla^2 \left[ -\frac{G}{c^2}\sum_a \frac{m_a}{r_a}-\frac{\Lambda}{6}|\Vec{x}|^2 \right] \,,
\end{eqnarray}
with $r_a:=|\Vec{x}-\Vec{x}_a(t)|$, and note that we have used the relations $\delta^{3}(\Vec{x}-\Vec{x}_a(t))=-\frac{1}{4\pi}\nabla^2 \left( \frac{1}{r_a} \right)$ and $\nabla^2 |\Vec{x}|^2=6$. Thus, the solution at lowest order is:
\begin{equation}\label{lowest_U}
U=-\frac{G}{c^2}\sum_a \frac{m_a}{r_a}-\frac{\Lambda}{6}|\Vec{x}|^2 \,.
\end{equation}
We remark that the substitution of \eqref{lowest_U} into the components of the metric \eqref{g_00} and \eqref{g_ij} leads to the same results previously given in \eqref{g_00_h} and \eqref{g_ij_h} using the DIRE approach. This, in fact, reflects that both approaches PN and PM are related to each other in the near zone of the faraway wave form. Furthermore, from \eqref{g_00_solution} and \eqref{g_ij_solution}, one can realize that the ansatz given by \eqref{g_00} and \eqref{g_ij} is satisfied providing that $U=-\frac{1}{2}\overset{(2)}{g}_{00}=-\frac{G}{c^2}\sum_a \frac{m_a}{r_a}-\frac{\Lambda}{6}|\vec{x}|^2$; hence, such ansatz matches to the solution of $\overset{(2)}{g}_{00}$ and $\overset{(2)}{g}_{ij}$.

Moreover, from \eqref{nabla_chi} and \eqref{nabla_xi_i}; and using the result \eqref{lowest_U}, we find that:
\begin{eqnarray}
    \label{solution_chi}\chi&=& -\frac{G}{2c^2}\sum_a m_a r_a+\frac{\Lambda}{24}|\Vec{x}|^4 \,, \\
    \label{solution_xi_i}\xi_i&=& -\frac{G}{c^3}\sum_a m_a \frac{v_{ai}}{r_a} \,.
\end{eqnarray}
This leads to: 
%
\begin{eqnarray}\label{g_i}
 && \hspace{-1.5cm} g_i = -\frac{G}{8c^3}\sum_a \frac{m_a}{r_a}\left[ 7v_{ai}+\hat{n}_{ai}(\hat{n}_a\cdot \hat{v}_a) 
 \right] \nonumber \\
 && \hspace{-1.0cm} + \frac{\Lambda}{24c}|\Vec{x}|^2 \left[ \frac{d x_i}{dt}+2\frac{x_i}{|\vec{x}|^2} (\vec{x}\cdot \Vec{v}) \right]-\frac{\Lambda}{6c}(x_i)^2\Dot{x}_i \,,
\end{eqnarray}
where $\hat{n}_a=\vec{r}_a/r_a$ and $\hat{v}_a=d\vec{r}_a/dt$. These results \eqref{lowest_U} and \eqref{g_i} do comply the `Coulomb-like' gauge \eqref{Coulomb_gauge} at order $O(c^{-3}, \Lambda c^{-1})$. Subsequently, using the previous result at Newtonian order, one can get the following result:
\begin{eqnarray}\label{U_1PN}
&& \hspace{-1.5cm} U = -\frac{G}{c^2} \sum_a \frac{m_a}{|\Vec{x}-\Vec{x}_a(t)|}\nonumber\\
    &&\hspace{-1cm} -\frac{G}{c^4}\sum_a \frac{m_a}{|\Vec{x}-\Vec{x}_a(t)|}\left( \frac{3}{2}v_a^2-G\sum_{b\neq a}\frac{m_b}{r_{ab}} \right) \nonumber\\
    &&\hspace{-1cm} -\frac{G\Lambda}{3c^2}\sum_a \frac{m_a |\Vec{x}_a(t)|^2}{|\Vec{x}-\Vec{x}_a(t)|} \nonumber \\
    &&\hspace{-1cm} -\frac{\Lambda}{6}|\Vec{x}|^2+\frac{\Lambda}{c^2}G\sum_a m_a r_a+O(c^{-4},\Lambda c^{-2}) \,,
\end{eqnarray}
here $\vec{r}_{ab}:=\vec{r}_a-\vec{r}_b$. On the other hand, considering a two body compact system, the two body Lagrangian can be obtained à la Droste-Fichtenholz, a technique which, at this order is equivalent to the Fokker Lagrangian~\cite{Luc_Blanchet}. To obtain the Lagrangian that describes the interaction of compact bodies; i.e., which are regarded as point particles, we begin computing the equations of motion of a particle of mass $m_1$ moving in the near zone which follows the geodesic equation. This action is given by:
\begin{eqnarray}
&& \hspace{-1.2cm} S:=\int dt L_{m_1}=-m_1c\int dt \bigg(-g_{\mu\nu}\frac{dx^{\mu}}{dt}\frac{dx^{\nu}}{dt}\bigg)^{1/2} \nonumber\\
&& \hspace{-1.0cm} = -m_1c^2 \int dt \bigg(  -g_{00}-2g_{0i}\frac{v_1^i}{c}-g_{ij}\frac{v_1^iv_1^j}{c^2} \bigg)^{1/2} \,,
\end{eqnarray}
where $L_{m_1}$ is the Lagrangian of the geodesic of the mass $m_1$ and $v_1^i$ means the velocity of the particle $1$. We expand the integrand given by such Lagrangian to $1$PN order as follows:
\begin{widetext}
    \begin{eqnarray}\label{lagrangian_geodesic}
        L_{m_1}&=&-m_1c^2e^{U}\bigg[ 1-8g_i\frac{v_1^i}{c}e^{-2U}-e^{-4U}e^{-\frac{\Lambda}{2}|\vec{x}|^2}\bigg( e^{-\frac{\Lambda}{2}x_1^2}\frac{v_{1x}^2}{c^2}+e^{-\frac{\Lambda}{2}y_1^2}\frac{v_{1y}^2}{c^2}+e^{-\frac{\Lambda}{2}z_1^2}\frac{v_{1z}^2}{c^2} \bigg) \bigg]^{1/2} \nonumber\\
        &=&-m_1c^2 \bigg[ 1-\frac{1}{2c^2}v_1^2+U -\frac{1}{8c^4}v_1^4 +\frac{3}{2c^2}v^2U+\frac{1}{2}U^2-\frac{4}{c}g_{i}v_1^i+\frac{\Lambda}{4c^2}v_1^2|\vec{x}_1|^2+\frac{\Lambda}{4c^2}\left( x_1^2v_{1x}^2+y_{1}^2v_{1y}^2+z_1^2v_{1z}^2 \right) \bigg] \nonumber\\
        && + O(c^{-4},\Lambda c^{-2}, \Lambda^2) \,,
    \end{eqnarray}
\end{widetext}
where we utilized the ansatz of the metric given in \eqref{g_00}, \eqref{g_0i}, and \eqref{g_ij}; and evaluate the Lagrangian at the position of the mass $m_1$, implying that we must assess the potentials $U$ and $g_i$ on the trajectory, where their self part $\propto m_1$ or $m_1^2$ diverges, ignoring all the contributions to the field from the body $m_1$. These ill-defined (formally infinite) potentials that diverge are regularized (see for instance~\cite{PhysRevD.51.5360}) yielding:
\begin{widetext}
\begin{eqnarray}
    \label{U_reg}U&=&-\frac{Gm_2}{c^2 r_{12}}\left( 1+\frac{3}{2c^2}v_2^2 \right) -\frac{G\Lambda m_2}{3c^2r_{12}}|\vec{x}_2|^2-\frac{\Lambda}{6}|\vec{x}_1|^2+\frac{\Lambda G}{c^2}m_2 r_{12} \,, \\
    \label{g_reg}g_{i}&=& -\frac{Gm_2}{8c^3r_{12}}\bigg[ 7v_{2i}+\hat{n}_{2i}(\hat{n}_{2}\cdot \vec{v}_2) \bigg]+\frac{\Lambda}{24c}|\vec{x}_1|^2\left[ v_{1i}+2(\hat{n}_1\cdot \vec{v}_1)\hat{n}_{1i} \right]-\frac{\Lambda}{6c}(x_{i1})^2 \dot{x}_{1i} \,.
\end{eqnarray}
Substituting the regularized potentials \eqref{U_reg} and \eqref{g_reg} into the Lagrangian \eqref{lagrangian_geodesic} leads to:
\begin{eqnarray}\label{L_m1}
    L_{m_1}&=&-m_1c^2+\frac{1}{2}m_1v_1^2+\frac{Gm_1m_2}{r_{12}}+\frac{1}{8c^2}m_1v_1^4+\frac{Gm_1m_2}{2c^2 r_{12}}\bigg[ 3(v_1^2+v_2^2)-7\vec{v}_1\cdot \vec{v}_2-(\hat{n}_{12}\cdot \vec{v}_1)(\hat{n}_{12}\cdot \vec{v}_2) \bigg]-\frac{G^2 m_1 m_2^2}{2c^2 r_{12}^2}\nonumber\\
    && + \frac{G\Lambda m_1m_2}{3r_{12}}\left( r_2^2-\frac{1}{2}r_1^2 \right)+\frac{\Lambda c^2 m_1 r_1^2}{6}-\Lambda Gm_1 m_2 r_{12}+\frac{\Lambda}{6}m_1r_1^2\left[ v_1^2+2(\hat{n}_1\cdot \vec{v}_1)^2 \right] \nonumber\\
    && - \frac{11}{12}\Lambda m_1 (x_1^2 v_{1x}^2+y_1^2v_{1y}^2+z_1^2v_{1z}^2)+O(c^{-4},\Lambda c^{-2}, \Lambda^2) \,.
\end{eqnarray}
\end{widetext}
The Fichtenholz Lagrangian that describes the motion of two compact bodies is constructed out in such a way to give the same equations of motion as \eqref{L_m1} when $m_1\rightarrow 0$. Hence, the Lagrangian that governs the motion of two compact bodies in interaction is given by:
\begin{widetext}
    \begin{eqnarray}\label{lagrangian_two_bodies}
    L&=&-m_1c^2-m_2c^2+\frac{1}{2}m_1v_1^2+\frac{1}{2}m_2v_2^2+\frac{Gm_1m_2}{r_{12}}+\frac{1}{8c^2}m_1v_1^4+\frac{1}{8c^2}m_2v_2^4 \nonumber\\
    && + \frac{Gm_1m_2}{2c^2 r_{12}}\bigg[3(v_1^2+v_2^2)-7\Vec{v}_1\cdot \Vec{v}_2-(\hat{n}_{12}\cdot \Vec{v}_1)(\hat{n}_{12}\cdot \Vec{v}_2)\bigg]-\frac{G^2m_1 m_2}{2c^2 r_{12}^2}(m_1+m_2) \nonumber \\
    && + \frac{\Lambda}{6}c^2(m_1r_1^2+m_2r_2^2)+G\Lambda m_1 m_2\left( \frac{r_1^2+r_2^2}{6r_{12}}-r_{12} \right)+\frac{\Lambda}{6}m_1 r_1^2 v_1^2+\frac{\Lambda}{6}m_2 r_2^2 v_2^2 \nonumber\\
    && + \frac{\Lambda}{3}m_1(\hat{n}_1\cdot \Vec{v}_1)^2r_1^2+\frac{\Lambda}{3}m_2(\hat{n}_2\cdot \Vec{v}_2)^2r_2^2 \nonumber \\
    && - \frac{11}{12}\Lambda \bigg[ m_1\left( x_1^2v_{1x}^2+y_1^2v_{1y}^2+z_1^2v_{1z}^2 \right)+ m_2\left( x_2^2v_{2x}^2+y_2^2v_{2y}^2+z_2^2v_{2z}^2 \right) \bigg]+O(c^{-4}, \Lambda c^{-2}, \Lambda^2) \,,
\end{eqnarray}
\end{widetext}
with $\hat{n}_1=\frac{\Vec{x}-\Vec{x}_1(t)}{|\Vec{x}-\Vec{x}_1(t)|}$, $\hat{n}_2=\frac{\Vec{x}-\Vec{x}_2(t)}{|\Vec{x}-\Vec{x}_2(t)|}$. The first two lines of the Lagrangian correspond to the case of a null $\Lambda$. Besides, this formula depends explicitly of the components of the position and the velocity of the particles. On the other hand, this computation can be repeated taking into account the interaction of $n$ particles, giving as a result:
\begin{widetext}
\begin{eqnarray}
    L&=&\sum_a m_av_a^2+\sum_{a\neq b}\frac{Gm_am_b}{2r_{ab}}+\sum_a \frac{1}{8}m_a v_a^4-\sum_{a\neq b}\frac{Gm_am_b}{4r_{ab}}\left[ 7\vec{v}_a\cdot \Vec{v}_b+(\hat{n}_{ab}\cdot \vec{v}_a)(\hat{n}_{ab}\cdot \vec{v}_{b}) \right] \nonumber\\
    && + \frac{3G}{2}\sum_a\sum_{b\neq a}\frac{m_am_bv_a^2}{r_{ab}} -\frac{G^2}{2}\sum_a\sum_{b\neq a}\sum_{c\neq a}\frac{m_am_bm_c}{r_{ab}r_{ac}} \nonumber\\
    && + \frac{\Lambda}{6}c^2 \sum_a m_a r_a^2+\frac{G\Lambda}{6}\sum_a \frac{m_am_b}{r_{ab}}r_a^2-\frac{G\Lambda}{2}\sum_{a\neq b}m_am_b r_{ab}+\frac{\Lambda}{6}\sum_a m_ar_a^2 v_a^2+\frac{\Lambda}{3}\sum_a m_a (\hat{n}_a\cdot \vec{v}_a)^2r_a^2 \nonumber\\
    && -\frac{11}{12}\Lambda \sum_a m_a \left( x_a^2v_{ax}^2+y_a^2v_{ay}^2+z_a^2v_{az}^2 \right)+O(c^{-4},c^{-2}\Lambda, \Lambda^2) \,,
\end{eqnarray}
\end{widetext}
where $a=1,\dots, N$ labels the particle, $r_{ab}$ is the distance between the particle $a$ and $b$, and $\hat{n}_{ab}$ is the unit vector from $a$ to $b$. Considering the center of mass frame given by \eqref{y_1} and \eqref{y_2}, the Lagrangian \eqref{lagrangian_two_bodies} becomes:
\begin{widetext}
    \begin{eqnarray}\label{two_body_lagrangian_CM}
    L&=& -mc^2 +\frac{1}{2}\mu v^2+\frac{G\mu m}{r}+\frac{1}{8c^2}\mu v^4 (1-3\nu)+\frac{G\mu m}{2c^2 r}\bigg[ (3+\nu)v^2 +\nu(\hat{n}\cdot \Vec{v})^2-\frac{Gm}{r}\bigg]+\frac{\Lambda}{6}c^2 \mu r^2 \nonumber \\
    && - \frac{1}{6}G\Lambda \mu r(5+2\nu)+\frac{1}{6}\Lambda \mu (1-3\nu)r^2 v^2+\frac{\Lambda}{3}\mu(\hat{n}\cdot \Vec{v})^2(1-3\nu)r^2 \nonumber \\
    && - \frac{11}{12}\Lambda\mu(1-3\nu)(x^2v_{x}^2+y^2v_{y}^2+z^2v_{z}^2) +O(c^{-4}, \Lambda c^{-2}, \Lambda^2) \,,
\end{eqnarray}
\end{widetext}
with $\vec{r}=\vec{x}_1-\vec{x}_2$,  $\hat{n}=\frac{\vec{x}_1-\vec{x}_2}{|\vec{x}_1-\vec{x}_2|}$, and $\vec{v}:=\vec{v}_1-\vec{v}_2$ are the relative vector distance, the relative unit vector and the relative velocity vector between the particles $1$ and $2$, correspondingly; and $x$, $y$, $z$ are the Cartesian components of the relative vector position $\vec{r}$ with $v_{x}$, $v_{y}$ and $v_{z}$ as their respective velocities components. The objects without the vector symbol stands only for the magnitude of the vector. The two body Lagrangian \eqref{two_body_lagrangian_CM} is one of the main results of this work. It describes the interaction of a two body compact system with relativistic correction considering the presence of the cosmological constant. Using the Euler-Lagrange equations:
\begin{equation}
    \frac{d}{dt}\left( \frac{\partial L}{\partial v^i} \right)-\frac{\partial L}{\partial r^i}=0 \,,
\end{equation}
the equation of motion of a two body compact system is displayed as follows:
\begin{widetext}
\begin{eqnarray}\label{EOM}
    a^i&=& -\frac{Gm}{r^2}\hat{n}^i+\frac{Gm}{c^2 r^2}\bigg\lbrace \bigg[ \frac{Gm}{r}(4+2\nu)-v^2(1+3\nu)+\frac{3}{2}\nu (\hat{n}\cdot \vec{v})^2 \bigg] \hat{n}^i+(4-2\nu)(\hat{n}\cdot \vec{v})v^i \bigg \rbrace+\frac{\Lambda}{3}c^2 r\hat{n}^i \nonumber \\
    && + \Lambda (1-3\nu)\bigg[ -\frac{5}{3}r(\hat{n}\cdot \vec{v})+\frac{11}{3}(r_i\dot{r}_{i})\bigg] v^i-Gm\Lambda \bigg[2(\frac{3}{4}+\nu)+\frac{11}{6}(1-3\nu)\frac{(r_i)^2}{r^2}\bigg] \hat{n}^i \nonumber \\
    && - \Lambda r (1-3\nu)\bigg[ \frac{1}{2}v^2+\frac{11}{6}(v_i)^2 \bigg] \hat{n}^i+O(c^{-4}, \Lambda c^{-2}, \Lambda^2) \,,
\end{eqnarray}
\end{widetext}
with no sum over the repeated index $i$; so this implies that is not possible to express the equations of motion using vector notation, unlike to the case where $\Lambda=0$. The energy of the system given by $\frac{\partial L}{\partial v^i}v^i-L$, that is:
\begin{widetext}
\begin{eqnarray}\label{energy}
    E&=& mc^2+\frac{1}{2}\mu v^2-\frac{G\mu m}{r}-\frac{\Lambda}{6}c^2 \mu r^2+\frac{1}{c^2}\bigg \lbrace \frac{3}{8}\mu (1-3\nu)v^4+\frac{G\mu m}{2r}\bigg[ (3+\nu)v^2 +\nu (\hat{n}\cdot \vec{v})^2+\frac{Gm}{r}\bigg]\bigg \rbrace \nonumber \\
    && + \frac{G\Lambda}{6}\mu mr(5+2\nu)+  \frac{1}{3}(1-3\nu)\Lambda \mu r^2 \bigg[\frac{1}{2}v^2+(\hat{n}\cdot \vec{v})^2 \bigg]-\frac{11}{12}\Lambda\mu (1-3\nu)(x^2 v_x^2+y^2 v_y^2+z^2 v_z^2) \,.
\end{eqnarray}
\end{widetext}
Both in the acceleration \eqref{EOM} and the energy \eqref{energy} hinges on explicitly of the Cartesian components of the relative position and velocity as a consequence that the cosmological constant breaks the rotational symmetry of the system. Next, taking the orbital plane coordinates:
\begin{eqnarray}
    \hat{n}&=& (\mathrm{cos}\phi, \mathrm{sin}\phi,0) \,, \\
    \hat{\phi}&=& (-\mathrm{sin}\phi, \mathrm{cos}\phi,0) \,, \\
    \hat{z}&=& (0,0,1) \,,
\end{eqnarray}
we have that the components of the acceleration can be written as:
\begin{equation}\label{acceleration_orbital}
    a^i=(\Ddot{r}-r\Dot{\phi}^2)\hat{n}^i+\frac{1}{r}\frac{d}{dt}(r^2 \dot{\phi})\phi^i \,.
\end{equation}
Therefore, comparing \eqref{EOM} and \eqref{acceleration_orbital}, and adding the three different equations for each component, given that $i=1,2,3$, we obtain:
\begin{widetext}
\begin{eqnarray}
    \label{r_eq}\Ddot{r}&=& r\dot{\phi}^2-\frac{Gm}{r^2}+\frac{\Lambda}{3}c^2 r+\frac{Gm}{c^2 r^2}\bigg[ \frac{1}{2}(6-7\nu)\dot{r}^2-(r\dot{\phi})^2 (1+3\nu)+\frac{2Gm}{r}(2+\nu)\bigg] \nonumber \\
    && - \frac{1}{3}\Lambda r(1-3\nu)\dot{r}^2 
    +\frac{1}{9}\Lambda r^3 (1-3\nu)\dot{\phi}^2-\frac{\Lambda}{18}Gm(38+3\nu)+O(c^{-4}, \Lambda c^2, \Lambda^2) \,, \\
    \frac{d}{dt}(r^2 \dot{\phi}^2)&=&2(2-\nu)\frac{Gm}{c^2}\dot{r}\dot{\phi}+2\Lambda(1-3\nu)r^3 \dot{r}\dot{\phi} \,.
\end{eqnarray}
\end{widetext}
\section{Radiated power formula}\label{Radiated_power_formula}
In this section we show that the radiated power formula does not contain $\Lambda$ provided that the condition $\Lambda h\rightarrow 0$ is satisfied. Thus, from~\eqref{source_LL_tensor_cosmological}, we can observe that the radiated power, taking into account $\Lambda$, reads:
\begin{eqnarray}\label{power_radiated_without_cosmological}
    P&=&c\int \bigg[ (-g)t_{\mathrm{LL}}^{0k}-2\frac{c^4}{16\pi G}\Lambda \mathfrak{g}^{-1/2}\mathfrak{g}^{0k} \bigg] dS_{k} \nonumber\\
    &=& c\int \bigg[ (-g)t_{\mathrm{LL}}^{0k}-\frac{c^4}{8\pi G}\Lambda \mathfrak{g}^{-1/2}\left( \eta^{0k}+h^{0k} \right) \bigg]dS_{k} \nonumber\\
    &=& c\int \bigg[ (-g)t_{\mathrm{LL}}^{0k}-\frac{c^4}{8\pi G}\mathfrak{g}^{-1/2}\left( \Lambda h^{0k} \right) \bigg]dS_{k} \nonumber\\
    &\simeq&c\int (-g)t_{\mathrm{LL}}^{0k}dS_{k} \,,
\end{eqnarray}
where $dS_{k}$ is an outward-directed surface element on the two-dimensional surface $S$. Considering the shortwave approximation (see for example~\cite{poisson2014gravity}), which is based on expansion of the gravitational potentials in powers of $\lambda/R\ll 1$, with $\lambda$ as the wavelength of the source and $R$ as the distance between the source and the observation point, we write:
\begin{equation}\label{h_in_powers_lambda}
    h^{\alpha \beta}=(\lambda/R)f_1^{\alpha\beta}+(\lambda/R)^2f_{2}^{\alpha\beta}+
    \cdots \,,
\end{equation}
where $f_{n}^{\alpha\beta}$ with $n=1,2,3,\dots ,$ is a function of the retarded time $\tau:=t-\frac{R}{c}$. Substituting \eqref{h_in_powers_lambda} in $t_{\mathrm{LL}}^{0k}$, given by \eqref{Landau_Lifshitz_tensor}; and from there, we replace it into \eqref{power_radiated_without_cosmological}, yielding:
\begin{equation}
    P=\frac{c^3 R^2}{32\pi G}\int \dot{h}_{\mathrm{TT}}^{ij}\dot{h}^{\mathrm{TT}}_{ij}d^3 x \,.
\end{equation}
Observe that the assumption $\Lambda h\rightarrow 0$, due to the very small value of $\Lambda$, implies that the flux of the radiation power $P$ does not contain $\Lambda$; so giving as a result the expression \eqref{flux_energy_h_ij}. Nonetheless, in the waveform \eqref{wave_form_complete} $\Lambda$ does appear explicitly.  
\section{Energy loss rate obtained from the symmetric trace free (STF) multipole decomposition}\label{STF}
It is well known that the EW multipoles are related with the symmetric trace free multipoles at $1$PN order as follows \cite{RevModPhys.52.299, PhysRevD.54.4813}:
\begin{eqnarray}
\label{I_STF_ij}I_{\mathrm{STF}}^{ij}&=&I_{\mathrm{EW}}^{<ij>} \nonumber\\
    && \hspace{-1cm} + \frac{1}{21}\bigg( 11I_{\mathrm{EW}}^{<ij>kk}-12I_{\mathrm{EW}}^{k<ij>k}+4I_{\mathrm{EW}}^{kk<ij>} \bigg) \,, \\
\label{I_STF_ijk}\dot{I}_{\mathrm{STF}}^{ijk}&=& 3I_{\mathrm{EW}}^{ijk} \,, \\
\label{J_STF_ij}J_{\mathrm{STF}}^{ij}&=& \frac{1}{2}\epsilon^{<i}{}_{kl}I_{\mathrm{EW}}^{j>kl} \,.
\end{eqnarray}
Then, recalling the results of the EW multipoles \eqref{two_index_moment}, \eqref{three_index_moment}, and \eqref{four_index_moment}, and considering the interaction of only two particles at the center of mass frame coordinates \eqref{y_1} and \eqref{y_2}, the STF moments \eqref{I_STF_ij}, \eqref{I_STF_ijk} and \eqref{J_STF_ij} become:
\begin{eqnarray}
\label{I_circ_ij}I_{\mathrm{STF}}^{ij}&=&\mu r^{<i}r^{j>} +\frac{\mu}{7c^2}(-5+8\nu)\frac{Gm}{r}r^{<i}r^{j>} \nonumber\\
&& + \frac{\mu}{c^2}\frac{29}{42}(1-3\nu)v^2r^{<i}r^{j>} \nonumber\\
&& + \frac{11}{21c^2}\mu(1-3\nu)r^2 v^{<i}v^{j>} \nonumber\\
&& - \frac{\Lambda \mu}{2}(1-3\nu)r^2r^{<i}r^{j>}, \\
\label{I_circ_ijk}\dot{I}_{\mathrm{STF}}^{ijk}&=&-3\frac{\mu \Delta m}{mc^2}v^{<i}r^{j}r^{k>} \,, \\
\label{J_circ_ij}J_{\mathrm{STF}}^{ij}&=&\frac{\mu \Delta m}{c^2 m}\epsilon^{kl<i}r^{j>}r_{k}v_{l} \,.
\end{eqnarray}
At $1$PN approximation, the radiated power in terms of the STF multipoles is given by~\cite{PhysRevD.54.4813, PhysRevD.51.5360}:
\begin{eqnarray}
\label{radiated_power_STF} P&=&-\frac{G}{c^5}\bigg\lbrace \frac{1}{5}\dddot{I}_{\mathrm{STF}}^{ij}\dddot{I}_{ij}^{\mathrm{STF}}+\frac{1}{c^2}\bigg[ \frac{1}{189}\overset{(4)}{I}{}_{ijk}^{\mathrm{STF}}\overset{(4)}{I}{}^{ijk}_{\mathrm{STF}} \nonumber\\
&&  + \frac{16}{45}\dddot{J}_{ij}^{\mathrm{STF}}\dddot{J}_{\mathrm{STF}}^{ij} \bigg]+O(c^{-4}, c^{-2}\Lambda, \Lambda^2) \bigg\rbrace \,.
\end{eqnarray}
Therefore, plugging back the results \eqref{I_circ_ij}, \eqref{I_circ_ijk}, and \eqref{J_circ_ij} into \eqref{radiated_power_STF} yields the energy loss rate of a circular motion of a binary compact system given by \eqref{radiated_power_1}. 
\section{Computation of the integral $I(\Theta)$}\label{integral_I_Theta}
In this appendix we compute the integral 
\begin{equation}
I(\Theta) = \int \frac{\sin\left(2\phi_{0\mathrm{PN}}\right)}{x^{13}}dx \,,
\end{equation}
considering the Newtonian phase (\ref{Newtonian_phase}) neglecting the additional $\Lambda$ term since $(\Lambda G^2 m^2)/c^4\ll 1$, i.e., we only take $\phi_{0\mathrm{PN}}\simeq -x^{-5}/(32\nu)$, and the Post-Newtonian parameter $x=\Theta^{-1/8}/2$, therefore: 
\begin{equation}
\phi_{0\mathrm{PN}}=-\frac{\Theta^{5/8}}{\nu} \,. 
\end{equation}
Then, we have: 
\begin{eqnarray}
&& I(\Theta) = \int \frac{\sin\left(2\phi_{0\mathrm{PN}}\right)}{x^{13}}dx \nonumber\\
&& = 512\int d\Theta \Theta^{1/2}\sin\left(\frac{2\Theta^{5/8}}{\nu}\right) = -\frac{1024}{125}\nu \Theta^{1/4}\times \nonumber\\
&& \left. \bigg[ 50\Theta^{5/8}\cos\left(\frac{2\Theta^{5/8}}{\nu}\right) - 35\nu\sin\left(\frac{2\Theta^{5/8}}{\nu}\right) \right.\nonumber\\
&& \hspace{0.1cm} 7i\nu\left(E_{\frac{3}{5}}\left(-\frac{2i\Theta^{5/8}}{\nu}\right)-E_{\frac{3}{5}}\left(\frac{2i\Theta^{5/8}}{\nu}\right)\right) \bigg] \,, \label{result_integral_I_Theta}
\end{eqnarray}
with $E_{n}(x):=x^{n-1}\Gamma(1-n,x)$ as the exponential integral function with $\Gamma(1-n,x)=\int_{x}^{\infty}t^{-n}e^{-t}dt$ as the incomplete Gamma function. On the other hand, we also compute the following expression:
\begin{eqnarray}
&& \hspace{-0.45cm} \int \Theta^{-11/8}I(\Theta)d\Theta = -\frac{8192\nu}{375}\Theta^{1/8}\left[ 35\nu\sin\left(\frac{2\Theta^{5/8}}{\nu}\right) \right.\nonumber\\
&& \left. + 20 \Theta^{5/8}\left(E_{\frac{1}{5}}\left(-\frac{2i\Theta^{5/8}}{\nu}\right) + E_{\frac{1}{5}}\left(\frac{2i\Theta^{5/8}}{\nu}\right)\right) \right.\nonumber\\
&& \left. + 7\nu\left(iE_{\frac{3}{5}}\left(\frac{2i\Theta^{5/8}}{\nu}\right) - iE_{\frac{3}{5}}\left(-\frac{2i\Theta^{5/8}}{\nu}\right)\right)\right] \,. \label{result_2_integral_I_Theta}   
\end{eqnarray}
%


	\bibliographystyle{apsrev4-1}
	
	\bibliography{Draft}
	
\end{document}